\documentstyle[11pt,epsfig,psfig]{article}
\newcommand{\AmS}{{\protect\the\textfont2
  A\kern-.1667em\lower.5ex\hbox{M}\kern-.125emS}}

\begin{document}
\title{\huge{Results from Bottomonia Production at the Tevatron 
and Prospects for the LHC}
\thanks{Research partially supported by CICYT under grant AEN99-0692}} 
\author{{\bf J.L. Domenech-Garret$^{a}$\thanks{domenech@evalo1.ific.uv.es}
$\ $and M.A. Sanchis-Lozano$^{b,c}$\thanks{Corresponding author: mas@evalo1.ific.uv.es}}
\\ \\
\it (a) Departamento de F\'{\i}sica At\'omica, Molecular y Nuclear \\
\it (b) Instituto de F\'{\i}sica
 Corpuscular (IFIC), Centro Mixto Universidad de Valencia-CSIC \\
\it (c) Departamento de F\'{\i}sica Te\'orica \\
\it Dr. Moliner 50, E-46100 Burjassot, Valencia (Spain) }
\maketitle 
\begin{abstract}
	We extend our previous analysis on inclusive heavy quarkonia 
        hadroproduction 
        to the whole  $\Upsilon(nS)$ (n=1,2,3) resonance family.
	We use a Monte Carlo framework with the colour-octet mechanism
        implemented in the PYTHIA event generator. We include in our 
        study higher order QCD
        effects such as initial-state emission of gluons and 
        Altarelli-Parisi evolution of final-state gluons. We extract some 
        NRQCD colour-octet matrix elements  relevant for $\Upsilon(nS)$ 
        (n=1,2,3) hadroproduction from CDF data at the Fermilab Tevatron. 
        Then we extrapolate to LHC energies to predict prompt 
        bottomonia production rates.
        Finally, we examine the prospect to probe the gluon density
        in protons from heavy quarkonia inclusive hadroproduction at high 
        transverse momentum and its feasibility in LHC general-purpose 
        experiments. 
\end{abstract}
\vspace{-15cm}
\large{
\begin{flushright}
  IFIC/00-87\\
  FTUV-00-1221\\
  \today
\end{flushright} }
\vspace{15cm}
\newpage
\section{Introduction}
Although the main goal of the LHC machine is the search for and the study
of the physics beyond the Standard Model, the expected huge rates of 
bottom quark production make especially interesting the foreseen
$B$ physics programme for the LHC project. In fact a specific experiment
(LHCb) will focus on B physics, while the two general-purpose
experiments ATLAS and CMS will dedicate special periods for
data taking to this aim (see for example Ref. \cite{tdr}). 
Among heavy flavour physics, heavy quarkonia production
and decays have historically played a very important role in 
the development and
test of Quantum Chromodynamics (QCD) as the best candidate to 
account for the strong interaction dynamics,
and likely will continue keeping an outstanding position in
this task.

Moreover, over the last decade hadroproduction of heavy quarkonia has 
received a lot of attention from both theoretical and experimental 
viewpoints, to explain the discrepancy between the so-called 
colour-singlet model (CSM) and the experimental data, amounting to a 
factor of about 50 for direct $J/\psi$ hadroproduction at the Tevatron. In 
particular, the colour-octet mechanism (COM) \cite{braaten} can
be viewed as the (relativistic) generalization of the CSM and hence
the most natural explanation for the unexpected surplus of heavy resonance
hadroproduction. Nevertheless, when applied to other production
processes like photoproduction at HERA, problems initially arose which
cast doubts on the validity of the COM, although
recent progress has been done allowing for a
better understanding of the situation \cite{wolf}. Furthermore, 
results from Tevatron on charmonia polarization (one of the
foremost predictions of the COM) seem to indicate
even the failure of a naive application of the colour
production mechanisms for charmonia \cite{kraemer}.

However, the COM  can be viewed as deriving from
a low energy effective theory, the Non Relativistic QCD
(NRQCD) \cite{bodwin}, so the question actually arising is whether 
NRQCD  is the correct
framework to deal with quarkonia production and decay. Perhaps the $v$
expansion does not converge well for charmonium and subleading
contributions cannot be neglected; perhaps
the heavy quark spin symmetry is broken to a larger extent than
expected. More work in this regard is required to clarify
the situation. On the other hand possibly NRQCD
is appropriate to describe bottomonia states and their production
because of the larger mass of the bottom quark. Hence checking the
COM in bottomonia hadroproduction is one of the challenges
of strong interaction physics over the next years.
Indeed there are alternative models in the
literature (based on QCD) trying to explain the
experimental facts (see for example \cite{baranov, hagler}). 
More astringent tests of heavy
quarkonia production are thus required to enlight the
situation, which can be qualified as rather confusing at present
\cite{kraemer}.

In a series of previous works (\cite{mas0,mas1,mas2,mas3,montp,mas5})
we have extensively analyzed charmonium 
hadroproduction in a Monte Carlo framework, using
PYTHIA 5.7 \cite{pythia,pythia2} event generator with the colour-octet model
implemented as a new routine in the generation code
\cite{mas2}.
Basically, such a production mechanism is based on the formation of
an intermediate coloured state during the hard partonic interaction, 
evolving non-perturbatively into
physical heavy resonances in the final state with certain
probabilities governed by NRQCD \cite{bodwin}.
In this work we extend our previous study of the 
$\Upsilon(1S)$ resonance \cite{mas4,mas00} to the
whole  $\Upsilon(nS)$ family below open bottom production, i.e.
$n=1,2,3$, using the CTEQ4L parton distribution function
(PDF). A similar analysis can be found in Ref. \cite{braaten2}
although limited to transverse momentum ($p_T$) values
higher than 8 GeV.

Although the discrepancies 
between the CSM and experimental cross sections on bottomonia hadroproduction
are smaller than those found for charmonia \cite{fermi}, still some
extra contribution should be invoked to account for the surplus observed 
at the Fermilab Tevatron. However,
we find that, analogously to the charmonium case
\cite{mas2}, those matrix elements (MEs) determined from Tevatron
data in other analyses \cite{cho} have to be lowered once
initial-state radiation of gluons is taken into account.
This is because of the raise of an ({\em effective}) intrinsic
momentum ($k_T$) of the interacting partons enhancing the 
moderate and high-$p_T$ tail
of the differential cross section for heavy quarkonia production
(for more details see Ref. \cite{mas2}).
This effect, as generated by the appropriate PYTHIA 
algorithm \cite{pythia,torn2}, is more pronounced - and 
likely more sound from a physical viewpoint - than a pure Gaussian 
smearing with a (required) large $<k_T>$ value. Besides PYTHIA,
in smoothing the production cross section,
endows us with the possibility of extending our analysis to
the small $p_T$ region of bottomonium production, keeping
the assumption on the validity of the cross section factorization.

The study of bottomonia production 
in hadron colliders should permit a stringent test
of the colour-octet production mechanism, particularly regarding
the predicted (mainly transverse) polarization of the resonance 
at high-$p_T$ \cite{braaten3}, whereas other approaches, like 
the colour evaporation model, predict no net polarization; 
indeed, LHC experiments 
will cover a wider range of transverse momentum than at the Tevatron,
allowing to explore the region $p_T^2>>4m_b^2$, where $m_b$ denotes
the bottom quark mass.

In this paper we also present the prospects to 
probe the gluon density of protons via heavy quarkonia inclusive
hadroproduction at high transverse momentum in the LHC. Our proposal
should be viewed along with
other related methods of constraining the gluon distribution in
hadrons like di-jet, lepton pair and prompt photon production
\cite{lai,berger,martin}. We
must clearly state that it relies on the dominance of a particular 
production mechanism at high $p_T$ (the COM) predicting a dominant
contribution from gluon fragmentation.
In spite of this and other assumptions (such as the validity of 
the factorization of the cross section), our feeling is that
LHC collaborations should keep an open mind
on all the possibilities offered by the machine, 
thereby exploring the feasibility of this proposal. 
For all these and other physical reasons, it is worth to estimate, as 
a first step, the foreseen
production rate of bottomonium resonances at the LHC and this
constitutes one of the goals of this work. 

We have based our analysis of bottomonia inclusive production
on the results from Run IB of the CDF collaboration 
\cite{fermi,fermi1,fermi2}
at the Fermilab Tevatron. This means significantly
more statistics than the data sample from Run IA, employed
in a former analysis \cite{cho}. However, the different sources
of prompt $\Upsilon(1S)$ production were not yet
separated along the full accessible
$p_T$-range, in contrast to charmonium production.
Hence we give in Section 2 the numerical values for some relevant 
combinations of long-distance MEs, including {\em direct} and {\em indirect} 
$\Upsilon(nS)$ inclusive production, extracted from the fit to the CDF 
experimental points. (Prompt resonance production includes 
both direct and indirect channels, the latter referring to 
feeddown from higher $\Upsilon(nS)$ and $\chi_{bJ}(nP)$ states.)
Nevertheless, we
still are able to estimate some colour-octet MEs for {\em direct} 
$\Upsilon(1S)$ production
from the measurements on different production sources at $p_T>8$ GeV
\cite{fermi2}. The extrapolation to LHC is shown in Section 3
where we show the predicted differential and integrated cross section
for all $\Upsilon(nS)$ resonances. In Section 4 we discuss 
heavy quarkonia inclusive hadroproduction as a probe of the gluon density in
protons. Finally, in the appendices at the end of the paper, we gather
those technical details and values of the parameters
employed in the generation.


\begin{figure}
\centerline{\hbox{                        
 \psfig{file=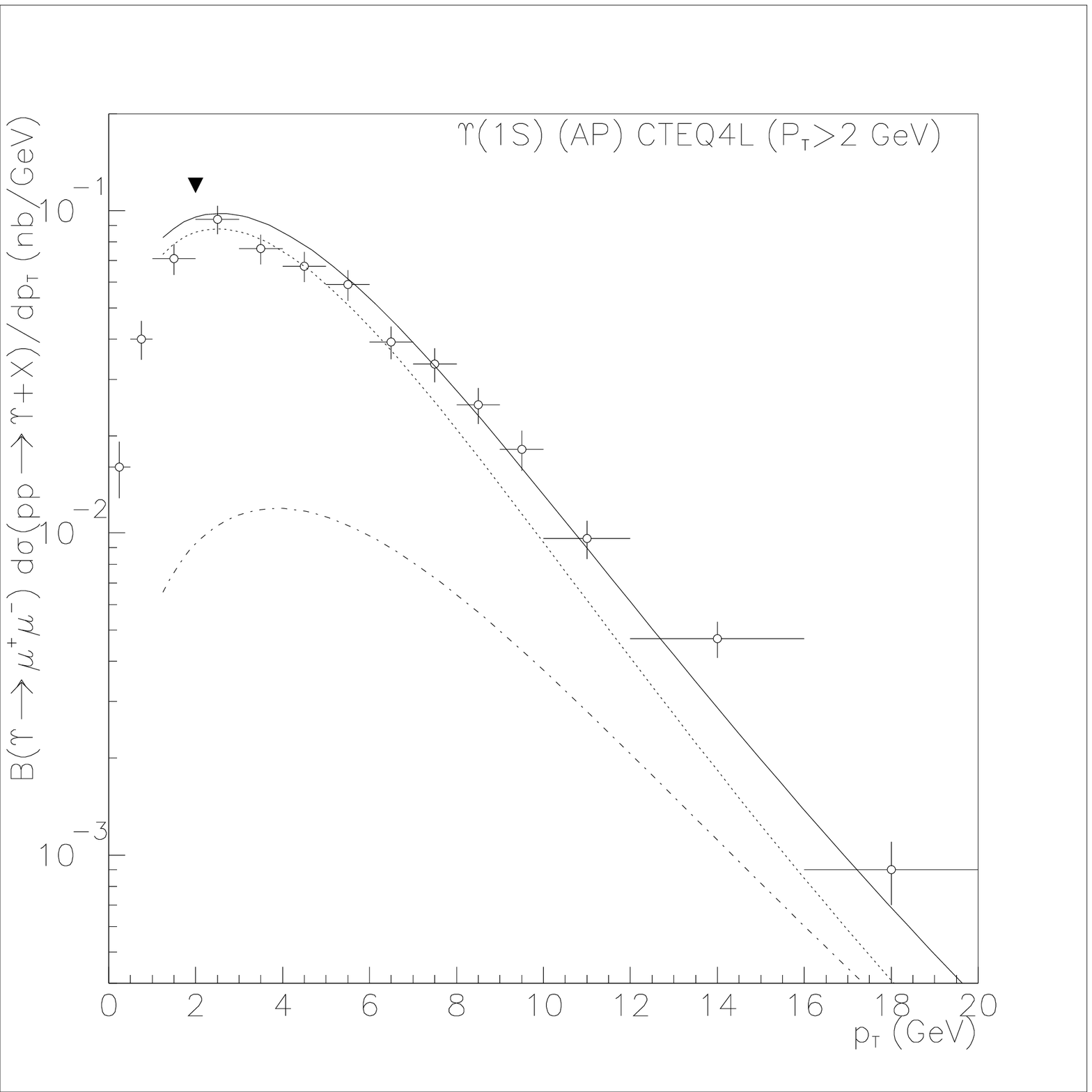,width=5.5cm} 
 \psfig{file=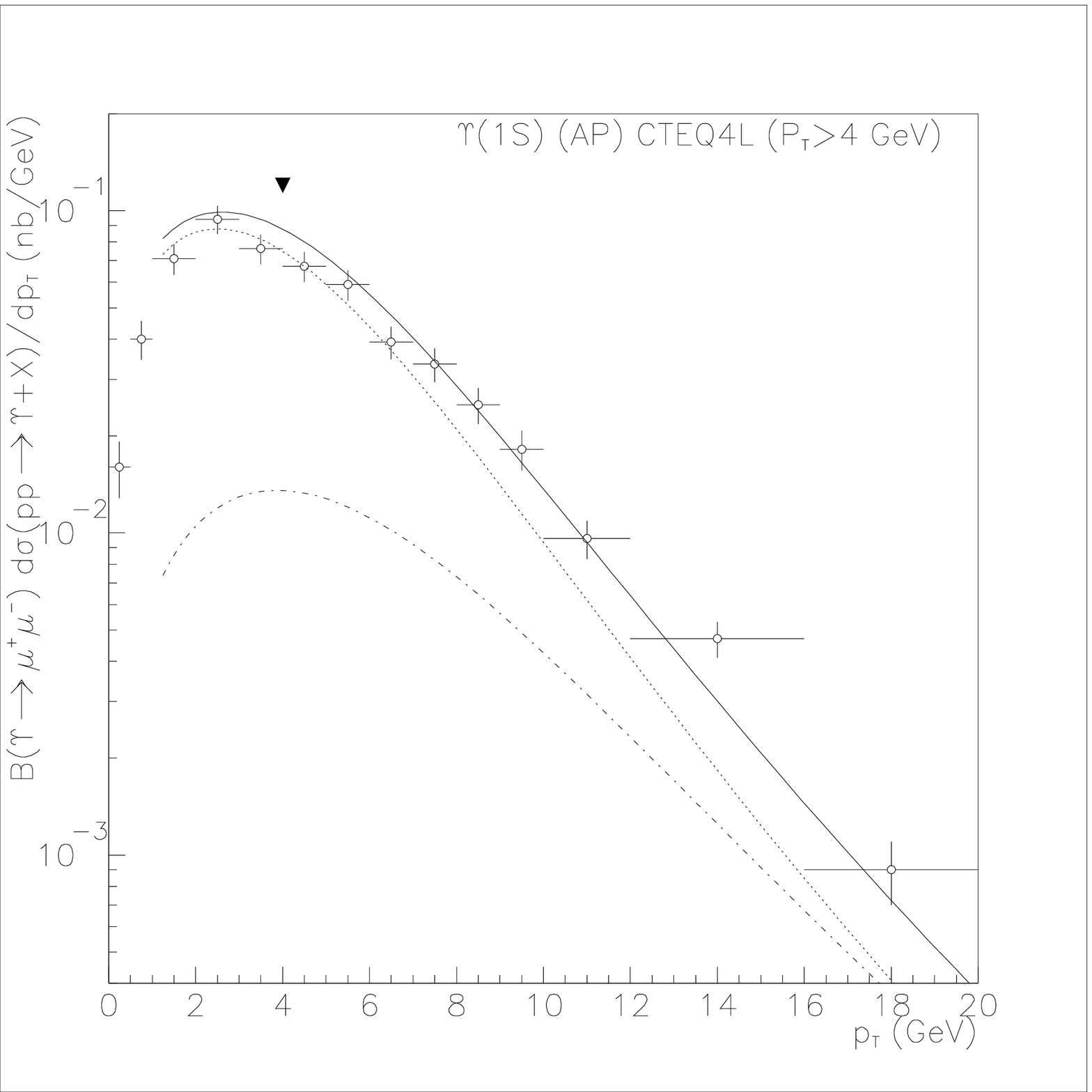,width=5.5cm} 
 \psfig{file=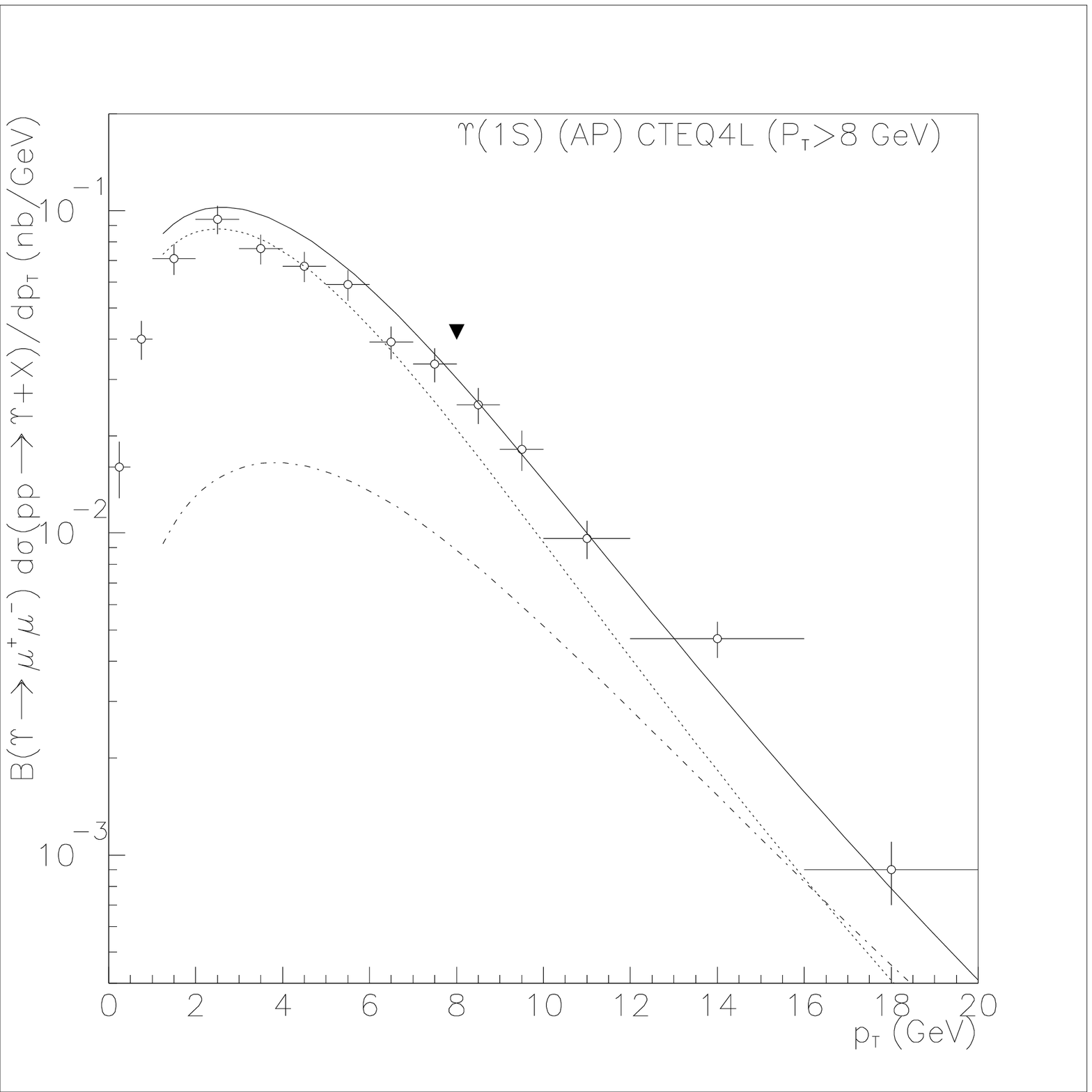,width=5.5cm}  
}}
\centerline{\hbox{
 \psfig{file=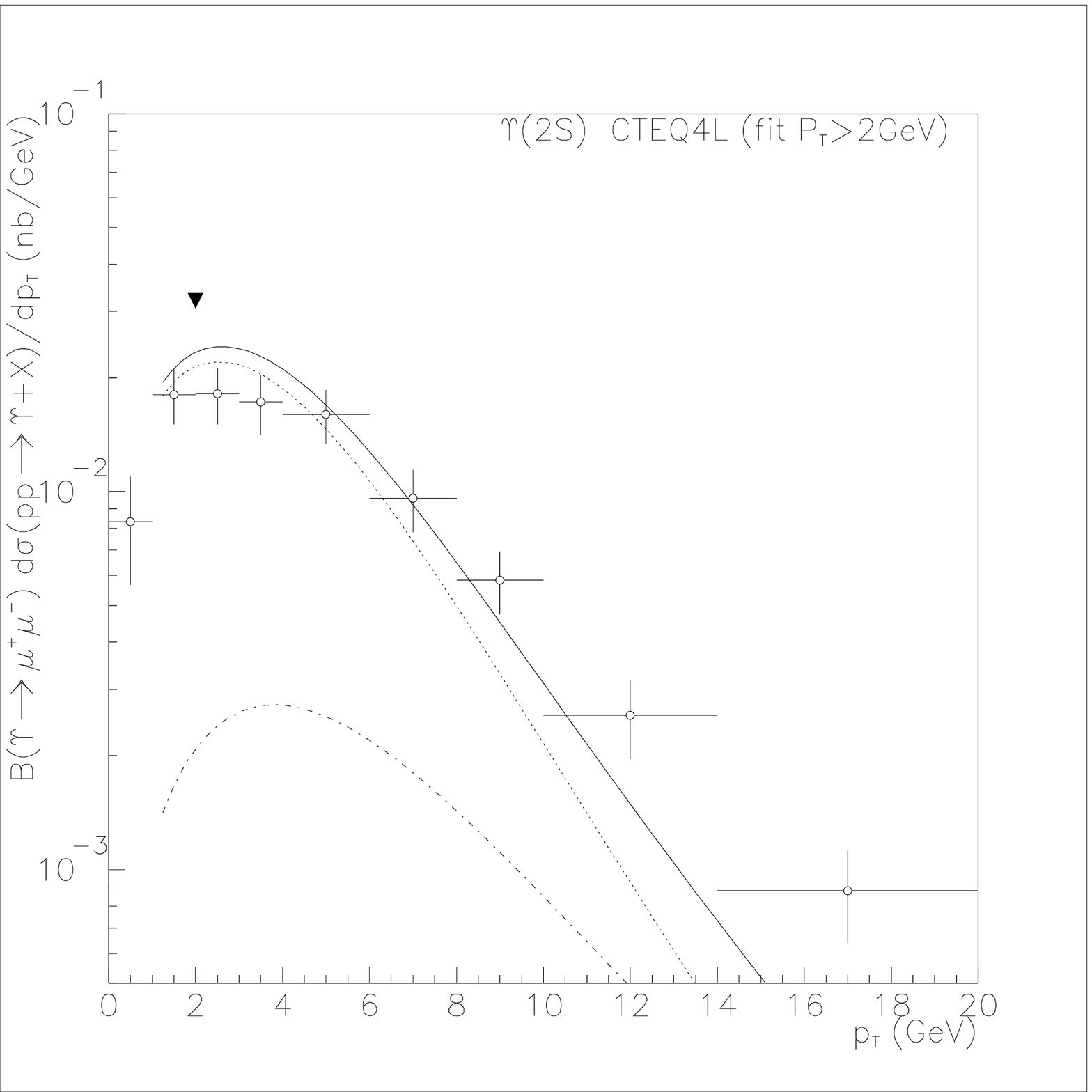,width=5.5cm} 
 \psfig{file=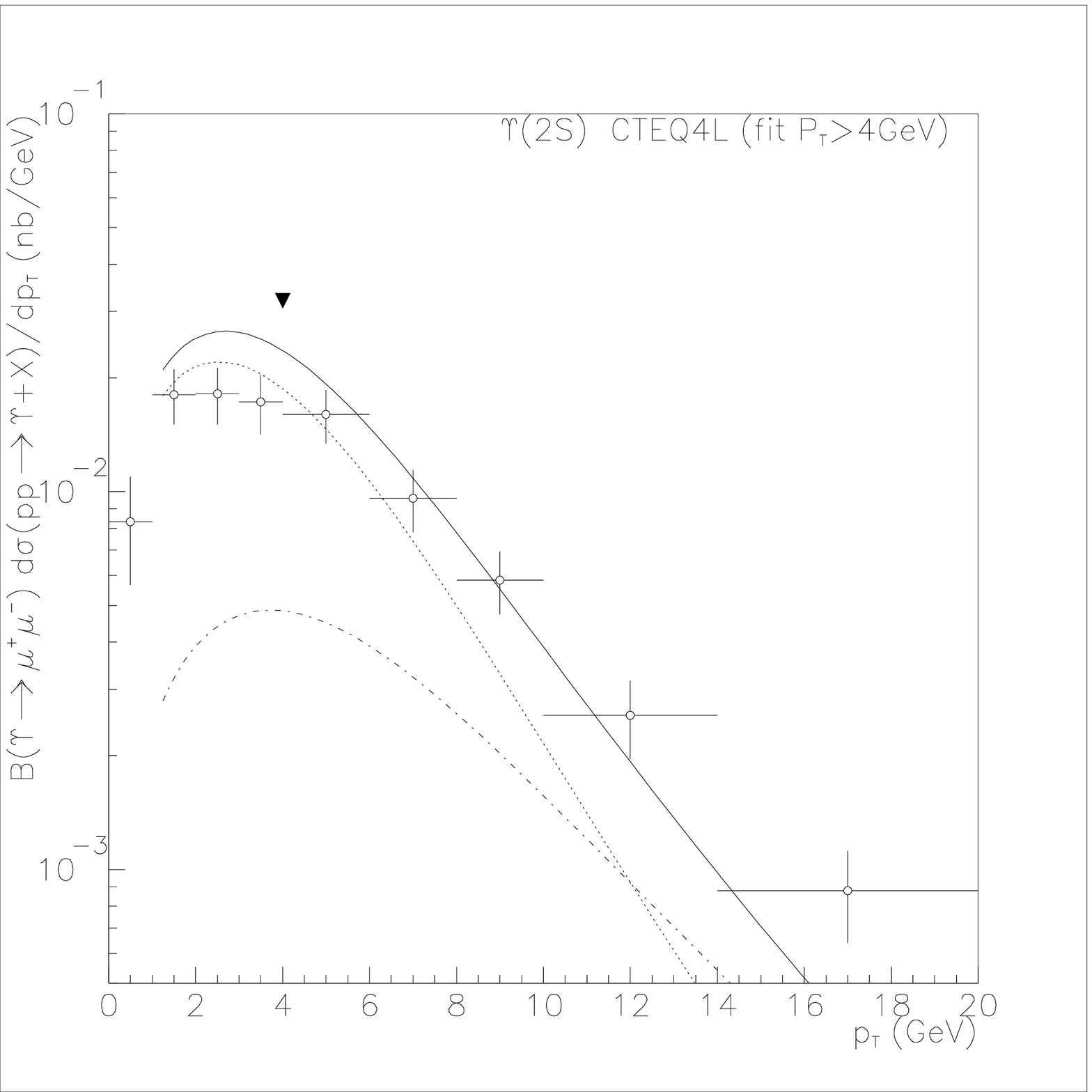,width=5.5cm} 
 \psfig{file=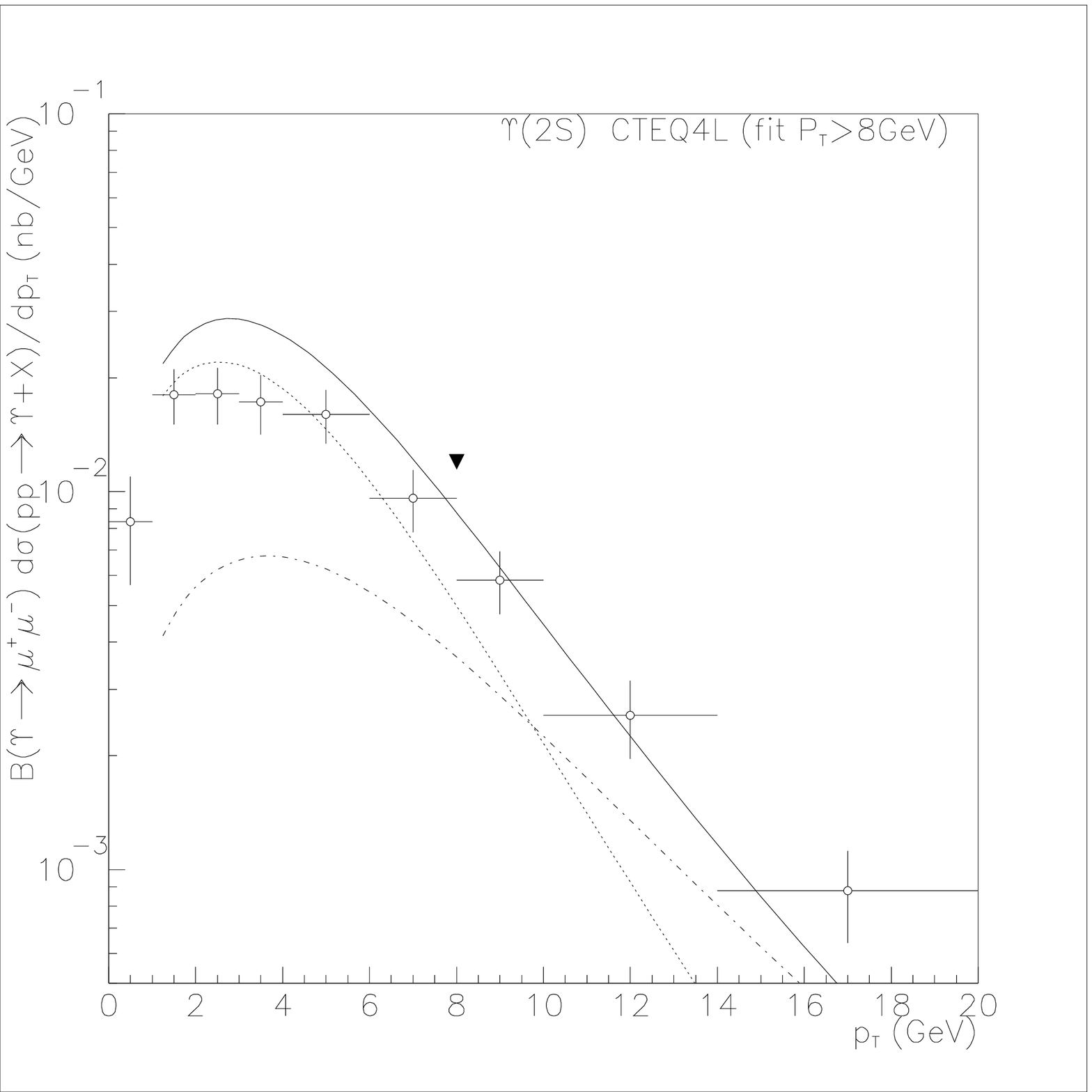,width=5.5cm}  
}}
\centerline{\hbox{ 
 \psfig{file=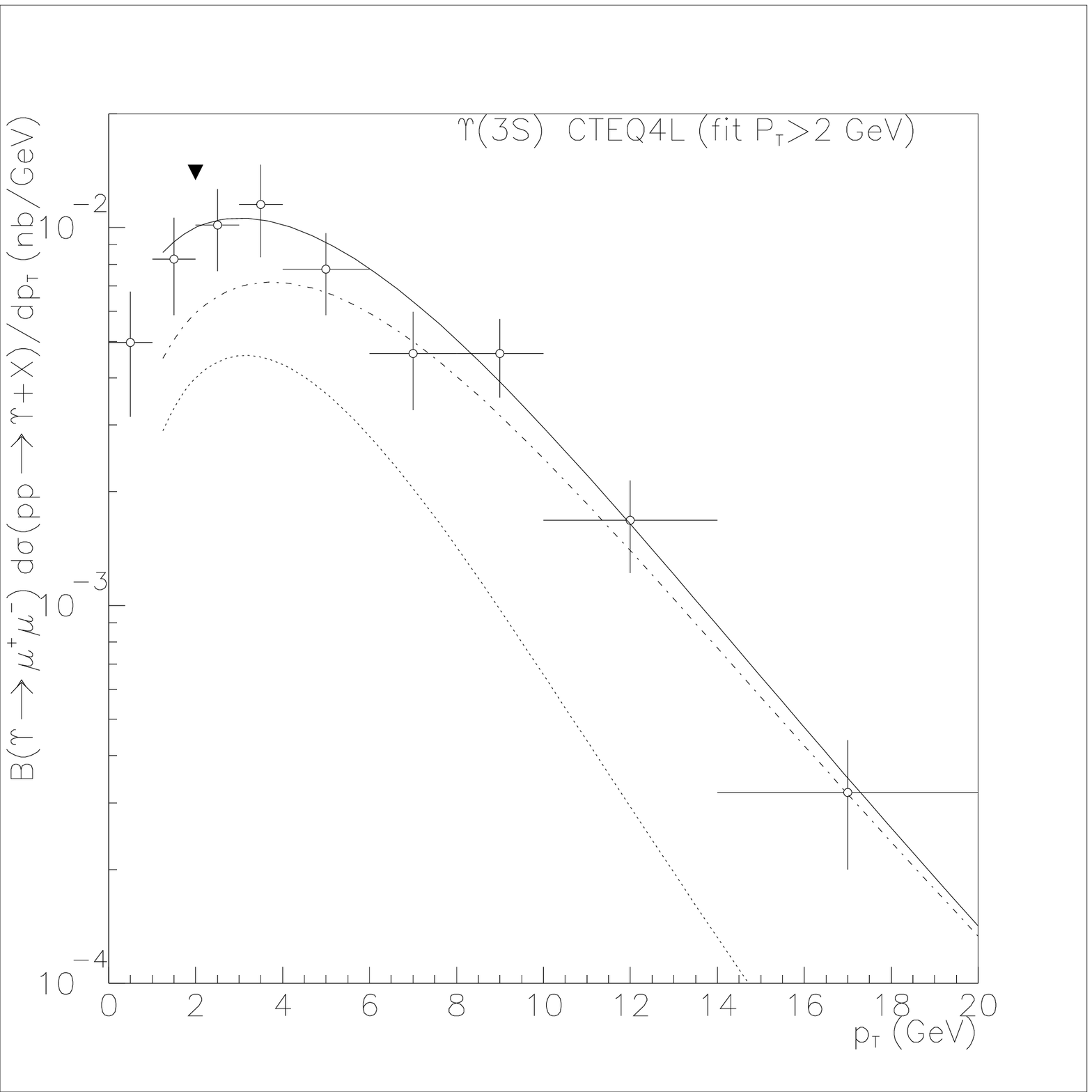,width=5.5cm} 
 \psfig{file=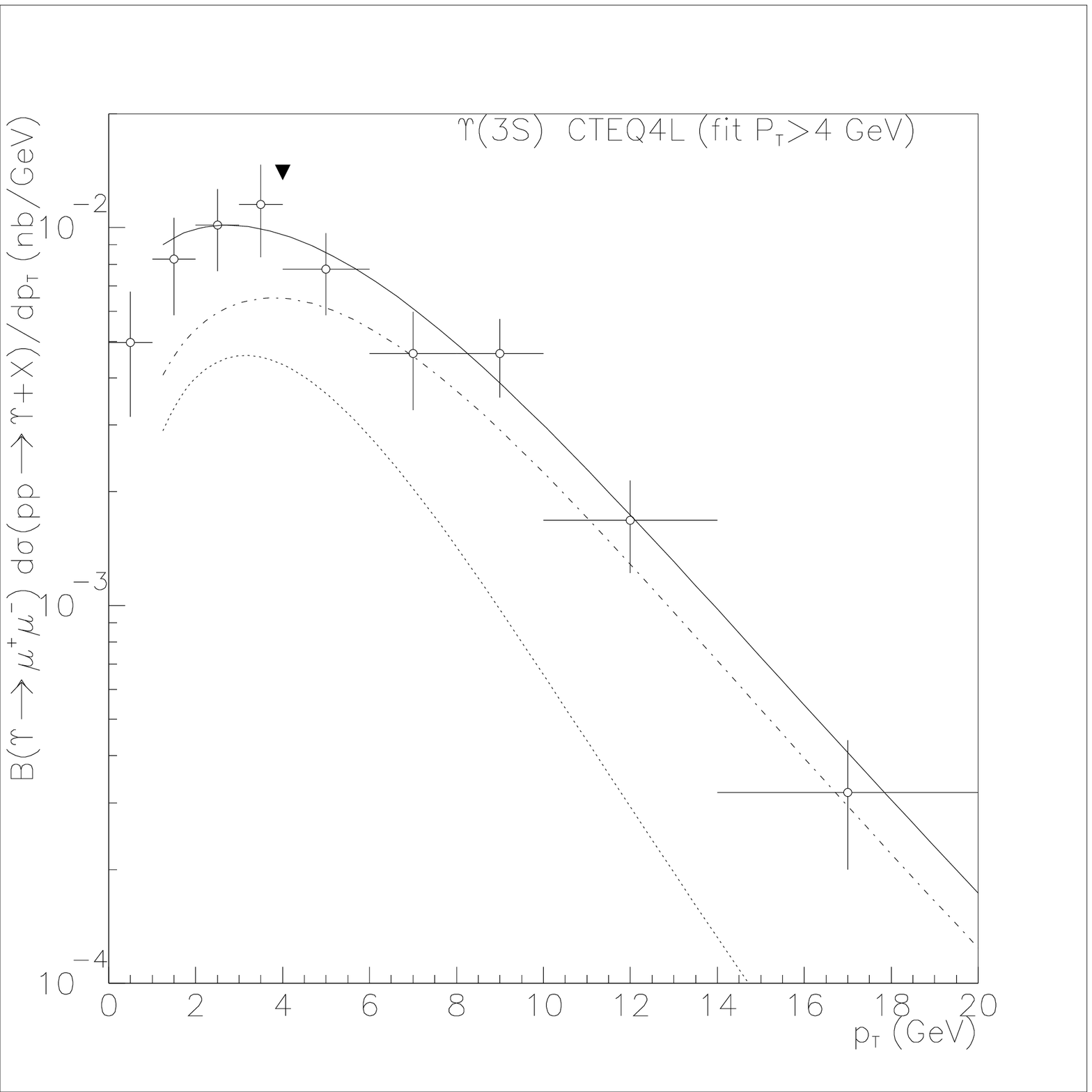,width=5.5cm} 
 \psfig{file=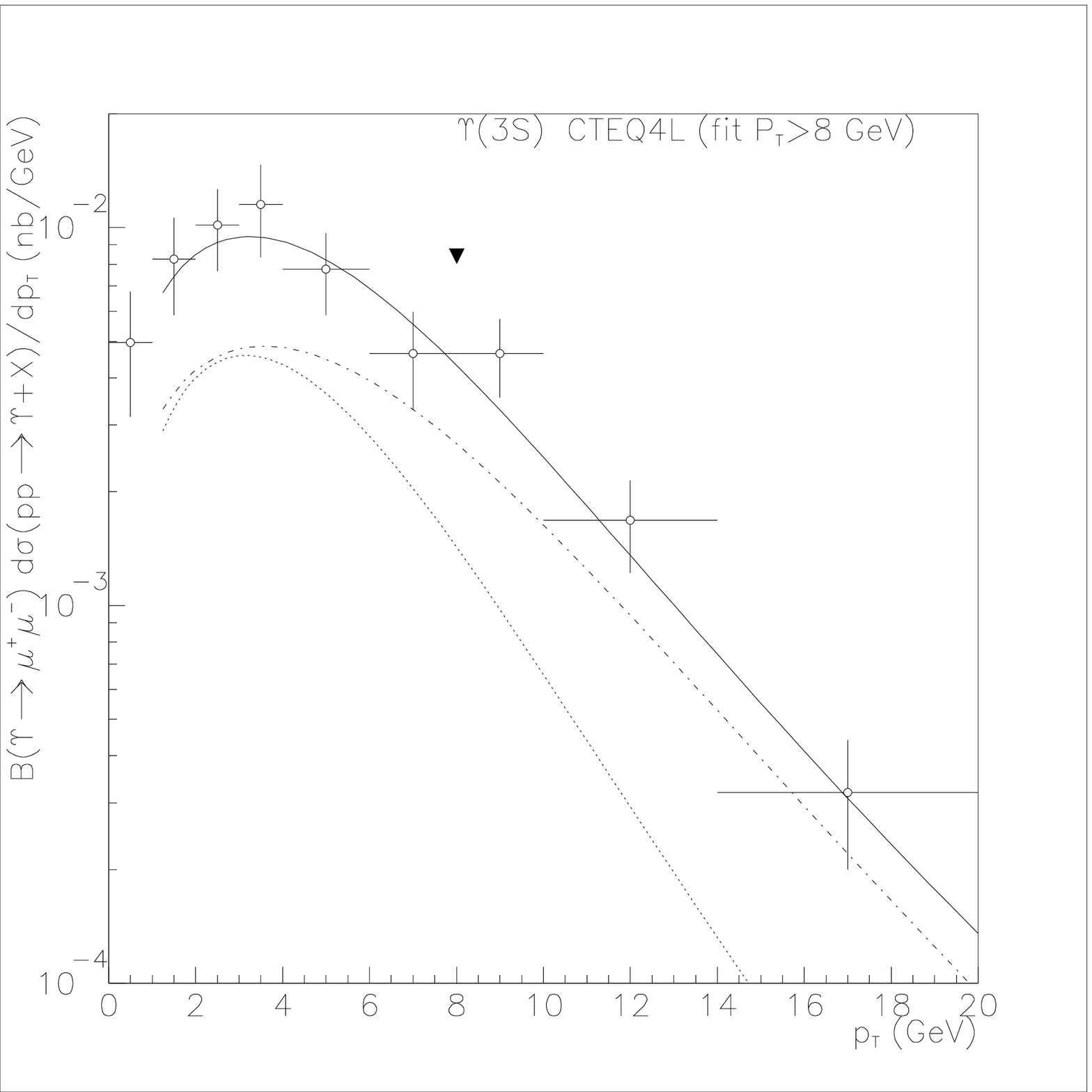,width=5.5cm}  
}}
\caption{Different fits to the Tevatron data on bottomonia 
inclusive production
in the rapidity interval ${\mid}y{\mid}<0.4$ using CTEQ4L PDF
and $m_b=4.88$ GeV.
{\em First row panels}: $\Upsilon(1S)$; 
{\em Second row panels}: $\Upsilon(2S)$;
{\em Third row panels}: $\Upsilon(3S)$. 
Dot, dot-dash and 
solid lines correspond to the CSM, COM ($^3S_1^{(8)}$ only) and all 
contributions, respectively. The triangle mark indicates
the $p_T$ lower cut-off used in the fit for each case: 2, 4 and 8 GeV. 
However, we plot the resulting curves extrapolating back over
$p_T>1$ GeV in all cases.}
\end{figure}

\section{Fits to Tevatron data}

As briefly outlined in the Introduction, the theoretical differential 
cross sections on inclusive production of bottomonia would stand 
above Tevatron experimental points for relatively high $p_T$ 
if the set of long-distance parameters from \cite{cho} were 
$\lq\lq$blindly'' employed in the PYTHIA generation running
with initial-state radiation on. This is the analogous conclusion 
which one of us (M.A.S.L.) reached to in the equivalent analysis 
performed on charmonia hadroproduction \cite{mas2}.
Indeed the smearing caused by multiple emission of gluons
by the interacting partons is not limited to small $p_T$
values as could be initially thought, but its influence spreads
over a larger region of transverse momenta. In fact we have
checked, from a fit to the $\Upsilon(3S)$ differential cross section, that 
actually this effect amounts to a pretty large value
for the effective $<k_T>$ of about 2 GeV, as we shall discuss
more extensively in Section 2.2.

Therefore we performed a new analysis of bottomonia
CDF data \cite{fermi1}, incorporating both 
direct and indirect production through the
CSM (as a $\lq\lq$fixed'' contribution which, in fact, is dominant
at low and even moderate $p_T$) and the COM, adjusting the long distance
parameters for different cut-offs from best 
${\chi}^2_{DF}{\equiv}{\chi}^2/N_{DF}$ fits to
the experimental points, using the CTEQ4L PDF. If not explicitly
stated the contrary, we are turning on initial-state radiation in
addition to a Gaussian primordial $k_T$ distribution 
(with $\sigma=0.44$ GeV, by default
in PYTHIA) in all generations.

\subsection{Extraction of the colour-octet  MEs}

In Figure 1 we show the theoretical curves obtained from our fits to
CDF data. In general, 
nice fits, with ${\chi}^2_{DF}$ values not too far from unity, were 
found, especially in the $\Upsilon(3S)$ case; instead, the $\Upsilon(2S)$
came out to be the worst one.
Let us stress that in the fitting procedure we excluded any 
possible {\em negative}
contribution from the different channels
at the cross section level, in contrast to \cite{braaten2}. Hence we 
had to dismiss
any contribution from the $^1S_0+^3P_J$ channels or, in other 
words, we set the $M_5$ long-distance parameter (as defined in 
appendix A) equal to zero, since any positive contribution from 
this channel would lead to a worse ${\chi}^2_{DF}$ value in 
all cases.

\begin{table*}[hbt]
\setlength{\tabcolsep}{1.5pc}
\caption{Values of $<O_8^{\Upsilon(nS)}(^3S_1)>{\mid}_{tot}$; $n=1,2,3$
(in units of $10^{-3}$ GeV$^3$) from 
the best fits to CDF data at the Tevatron on prompt 
$\Upsilon(nS)$ inclusive production for different $p_T$ lower cuts.
We also provide the ${\chi}_{DF}^2$ value for each case.
The CTEQ4L PDF was used with initial-state and AP
evolution activated in PYTHIA.}
\label{FACTORES}

\begin{center}
\begin{tabular}{ccccccc}    \hline
$p_T$ cut-off: & $2$ GeV  & ${\chi}_{DF}^2$ & $4$ GeV  & ${\chi}_{DF}^2$ 
& $8$ GeV  & ${\chi}_{DF}^2$  \\
\hline
$1S$ & $77{\pm}17$ & $1.74$ & $87{\pm}16$ & 1.53 & $106{\pm}13$ & 1.00 \\
\hline
$2S$ & $40{\pm}29$ & $2.87$ & $73{\pm}18$ & 1.58 & $103{\pm}27$ & 1.87 \\
\hline
$3S$ & $99{\pm}11$ & $1.00$ & $91{\pm}15$ & 1.00 & $68{\pm}11$ & 1.00 \\
\hline
\end{tabular}
\end{center}
\end{table*}

In Table 1 we show the values of $<O_8^{\Upsilon(nS)}(^3S_1)>{\mid}_{tot}$
($n=1,2,3$), as defined in (A.7), for different $p_T$ lower cut-offs, in 
correspondence with the plots of Figure 1. All values are roughly of order 
$10^{-1}$ GeV$^3$ and agree, within the errors, 
with the results obtained for $p_T>8$ GeV by the authors of 
Ref. \cite{braaten2} using the CTEQ5L parton distribution function.

Nevertheless, let us stress that our numerical estimates for the
colour-octet MEs have to be viewed with 
some caution because of the theoretical and $\lq\lq$technical'' 
(due to the Monte Carlo assumptions) uncertainties. For example, our 
algorithm for AP evolution (see appendix B) should be regarded as a 
way of reasonably steepening  the high-$p_T$ tail of the 
(leading-order) differential cross section, which otherwise would 
fall off too slowly as a function of $p_T$.

\subsubsection{Separated production sources for $p_T>8$ GeV}

Current statistics does not permit
to subtract indirect production sources 
to obtain the direct $\Upsilon(1S)$ production cross section
along the full accessible $p_T$-range. Nevertheless, 
feeddown from higher states ($\Upsilon(nS)$, ${\chi}_{bJ}(nP)$)
was experimentally separated out for $p_T>8$ GeV \cite{fermi,fermi2}.
We used this information to check our analysis {\em a posteriori}
(rather than using it as a constraint in the generation)
and to draw some important physical conclusions. To this end
the relative fractions of the contributing  channels
for $p_T>8$ GeV are reproduced  in Table 2 from Ref. \cite{fermi,fermi2}. 
On the other hand, we show in Table 3 (which updates our
results presented in Ref. \cite{mas4} using CTEQ2L) 
the fractions found in this work corresponding to the
different generated channels for  $p_T>8$ GeV, following the notation
introduced in appendix A.

\begin{table*}[hbt]
\setlength{\tabcolsep}{1.5pc}
\caption{Relative fractions (in $\%$) of the different contributions to 
$\Upsilon(1S)$ production from 
CDF data at $p_T>8$ GeV \cite{fermi2}. Statistical and
systematic errors have been summed quadratically.}

\label{FACTORES}

\begin{center}
\begin{tabular}{lcc}    \hline
contribution & Tevatron results \\ 
\hline
direct $\Upsilon(1S)$ & $50.9{\pm}12.2$ \\
\hline
$\Upsilon(2S)$+$\Upsilon(3S)$ & $11.5{\pm}9.1$ \\
\hline
${\chi}_b(1P)$ & $27.1{\pm}8.2$ \\
\hline
${\chi}_b(2P)$ & $10.5{\pm}4.6$ \\
\hline
\end{tabular}
\end{center}
\end{table*}

\begin{table*}[hbt]
\setlength{\tabcolsep}{1.5pc}
\caption{Relative fractions (in $\%$) of the different contributions 
to $\Upsilon(1S)$ production at the Tevatron for $p_T>8$ GeV from our 
generation. Possible contributions from
$\chi_{bJ}(3P)$ states were not generated.}

\label{FACTORES}

\begin{center}
\begin{tabular}{lcc}    \hline
contribution & our generation \\
\hline
$\Upsilon(1S){\mid}_{^3S_1^{(8)}}$ & $36.8$  \\
\hline
$\Upsilon(1S){\mid}_{CSM}$ & $19.5$  \\
\hline
$\Upsilon(2S)$+$\Upsilon(3S){\mid}_{CSM}$ & $3.9$   \\
\hline
${\chi}_b(1P){\mid}_{CSM}$ & $24.1$   \\
\hline
${\chi}_b(2P){\mid}_{CSM}$ & $15.7$  \\
\hline
\end{tabular}
\end{center}
\end{table*}

By comparison between Tables 2 and 3 we can conclude that
the $\Upsilon(1S)$ indirect production from $\chi_{bJ}$'s decays 
is almost completely accounted for by the CSM according to the
assumptions and values of the parameters presented in 
appendix A. Indeed, experimentally $37.6{\pm}9.4\%$ of 
$\Upsilon(1S)$ production is due to $\chi_{bJ}(1P)$ and
$\chi_{bJ}(2P)$ decays  \cite{fermi2}
while from our generation we find a similar global value, namely $39.8\%$, 
coming exclusively from colour-singlet production!   
Moreover, assuming that a $7.6\%$ from the $36.8\%$ fraction
(corresponding to  the colour-octet $^3S_1^{(8)}$ contribution 
as expressed in Eq. (A.7)) can be attributed to the 
$\Upsilon(2S)+\Upsilon(3S)$ channel in addition to the colour-singlet
contribution ($3.9\%)$, we obviously get the fraction $11.5\%$ 
for the latter,  
bringing our theoretical result into agreement with the
experimental value. This single assignment implies to 
reproduce quite well the experimental fraction 
(${\approx}\ 51\%$) of direct $\Upsilon(1S)$ production by
adding the remaining $^3S_1^{(8)}$ contribution to  the
${\Upsilon(1S)}_{CSM}$
channels (${\approx}\ 49\%$) in our generation.

Of course all the above counting was based on mean values from 
Table 2 and subject to rather large uncertainties. Nevertheless, apart from
the consistency of our generation w.r.t. experimental results
under minimal assumptions, we can conclude again as in \cite{mas4} 
that there is almost {\em no need for} $\Upsilon(1S)$ indirect production
from feeddown of $\chi_{bJ}$ states produced through
{\em the colour-octet mechanism}. In other words,
the relative contribution from $P$-wave states to 
$<O_8^{\Upsilon(1S)}(^3S_1)>{\mid}_{tot}$ in Eq. (A.7) should be 
quite smaller than
na\"{\i}vely expected from NRQCD scaling rules compared to the
charmonium sector, in agreement with 
some  remarks made in \cite{schuler} and recent results found in 
\cite{braaten2}.
The underlying reason for this discrepancy  w.r.t. other
analyses \cite{cho} can be traced back to
the dominant colour-singlet contribution to the cross section
at $p_T$ values as much large as ${\simeq}\ 18$ GeV (see Figure 1)
caused by the effective $k_T$ smearing - already applied to
charmonium hadroproduction by one of us \cite{mas2}.

On the other hand the corresponding velocity scaling rule
in the bottomonium sector is roughly verified as we shall see. Defining
the ratios of matrix elements:

\begin{equation}
R_v(n)\ =\ \frac{<O_8^{\Upsilon(nS)}(^3S_1)>{\mid}_{tot}}
{<O_1^{\Upsilon(nS)}(^3S_1)>{\mid}_{tot}}\ ,
\end{equation}
its values, shown in Table 4,
are in accordance with the expected order-of-magnitude 
$v^4\ {\approx}\ 0.01$, where $v$ is the relative velocity
of the bottom  quark inside bottomonium.
Nevertheless we realize an increase of $R_v(n)$ for higher
$n$ values. Assuming that the $<O_8^{\Upsilon(nS)}(^3S_1)>{\mid}_{tot}$ 
matrix element could be interpreted as a (weighted) colour-octet
wave function squared (in the
same way as $<O_1^{\Upsilon(nS)}(^3S_1)>{\mid}_{tot}$ w.r.t. the
colour-singlet state, see appendix A) the
ratio  $R_v(n)$ of both squared wave functions in the origin
comes out as not independent of the resonance state under consideration.

\begin{table*}[hbt]
\setlength{\tabcolsep}{1.5pc}
\caption{Values (in units of GeV$^3$) of different colour-singlet
and colour-octet combinations of MEs according to
Eqs. (A.4) and (A.7) and the ratios $R_v(n)$; $n=1,2,3$.
The best $\chi^2_{DF}$ values from Table 1 are displayed.}
\label{FACTORES}

\begin{center}
\begin{tabular}{lccc}    \hline
$Resonance $ & $<O_1^{\Upsilon(nS)}(^3S_1)>{\mid}_{tot}$ & 
$<O_8^{\Upsilon(nS)}(^3S_1)>{\mid}_{tot}$  & $R_v(n)$  \\ 
\hline
$\Upsilon(1S)$ & $11.1$ & $0.106$ & 0.0095   \\
\hline
$\Upsilon(2S)$ & $5.01$ & $0.073$ & 0.0145   \\
\hline
$\Upsilon(3S)$ & $3.54$ & $0.099$ & $0.028$   \\
\hline
\end{tabular}
\end{center}
\end{table*}

Thus we conclude that this particular NRQCD velocity scaling rule, although 
valid as an order-of-magnitude estimate, retains a weak dependence on 
the principal quantum number $n$, not completely 
cancelling in the ratio (1).

\begin{figure}[htb]
\centerline{\hbox{
\psfig{figure=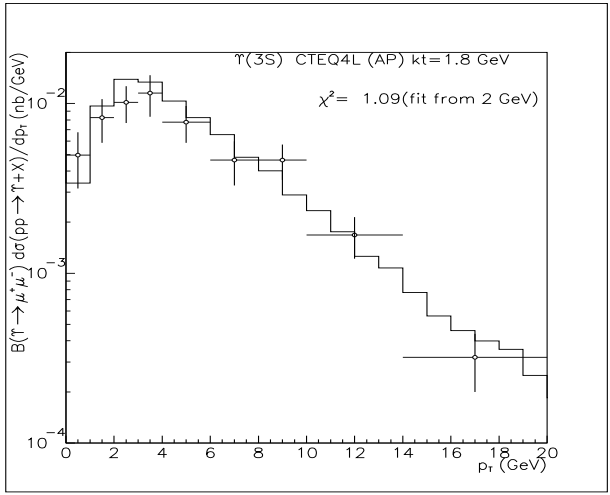,height=6.5cm,width=8.cm}
}}
\caption{Fit to the Tevatron data on $\Upsilon(3S)$ 
hadroproduction using a gaussian smearing
function with $\sigma=2$ GeV, i.e. $<k_T>=1.8$ GeV. }
\end{figure}

\subsection{Gaussian $<k_T>$ smearing}

The smearing effect on the differential cross section caused
by initial-state radiation of gluons can be roughly simulated by means of 
a gaussian intrinsic $k_T$ distribution of the interacting partons
inside hadrons, to be convoluted with the corresponding hard
interaction cross sections:
\begin{equation}
D({\bf k_T})\ =\ \frac{1}{{\pi}{\sigma}^2}
\exp{\biggl(-\frac{k_T^2}{\sigma^2}\biggr)}
\end{equation}
with
\begin{equation}
<k_T>\ =\ \frac{\sqrt{\pi}}{2}\ \sigma
\end{equation}

The width of the gaussian can be viewed as an adjustable
parameter \cite{sridhar}. In fact PYTHIA incorporates 
as an option a gaussian primordial
$k_T$ smearing, whose width can be
set by the user. We used this possibility
to make a $\lq\lq$new'' fit of Tevatron data for the
$\Upsilon(3S)$ resonance, employing the same matrix elements as
shown in Table 1 but with initial-state radiation off. 
Then the gaussian $k_T$ smearing has to simulate the
(this time missing) initial-state radiation. In 
Figure 2  we show the resulting histogram, corresponding to 
a value $\sigma=2$ GeV, i.e. $<k_T>=1.8$ GeV.

\begin{figure}
\centerline{\hbox{
\psfig{figure=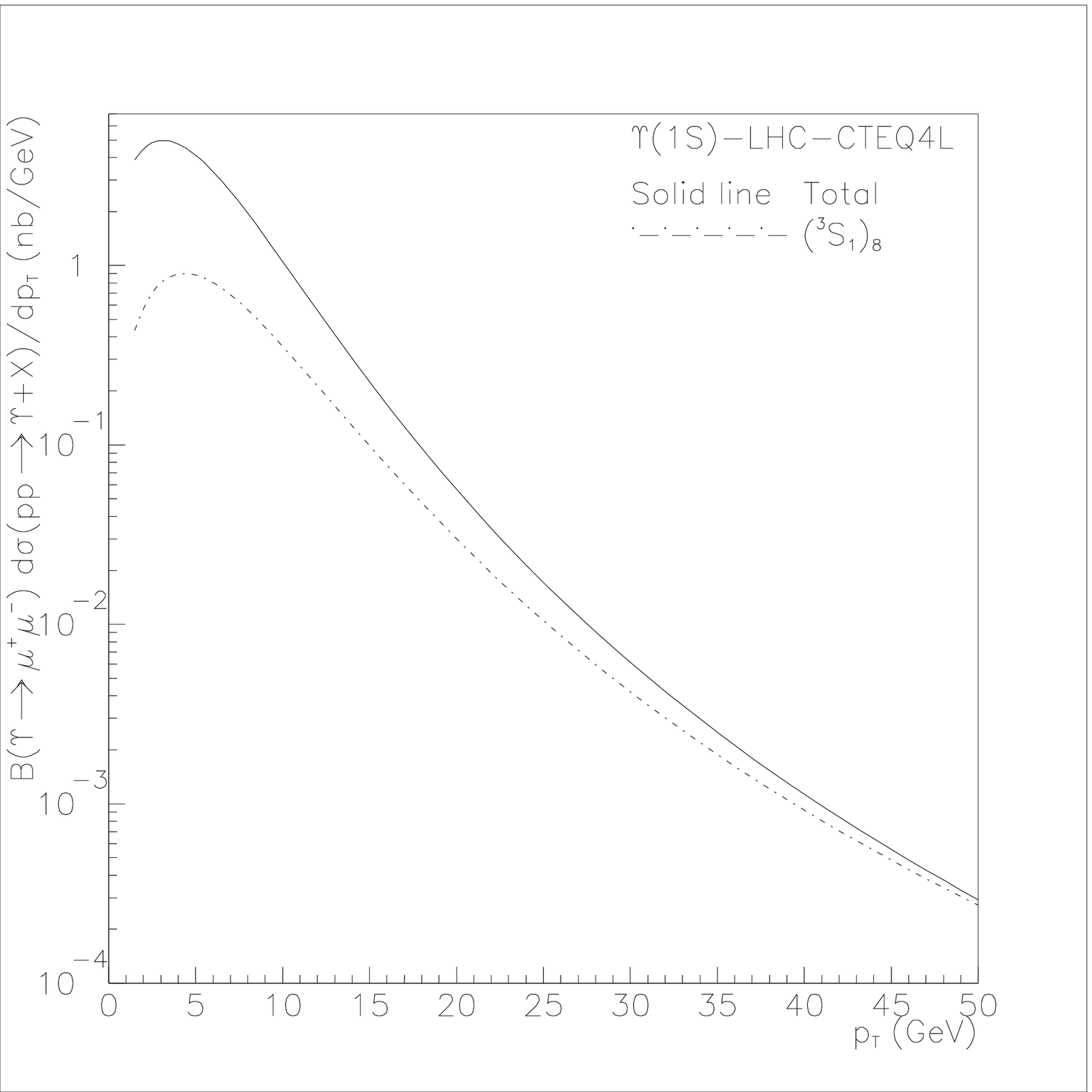,height=6.5cm,width=8.cm}
\psfig{figure=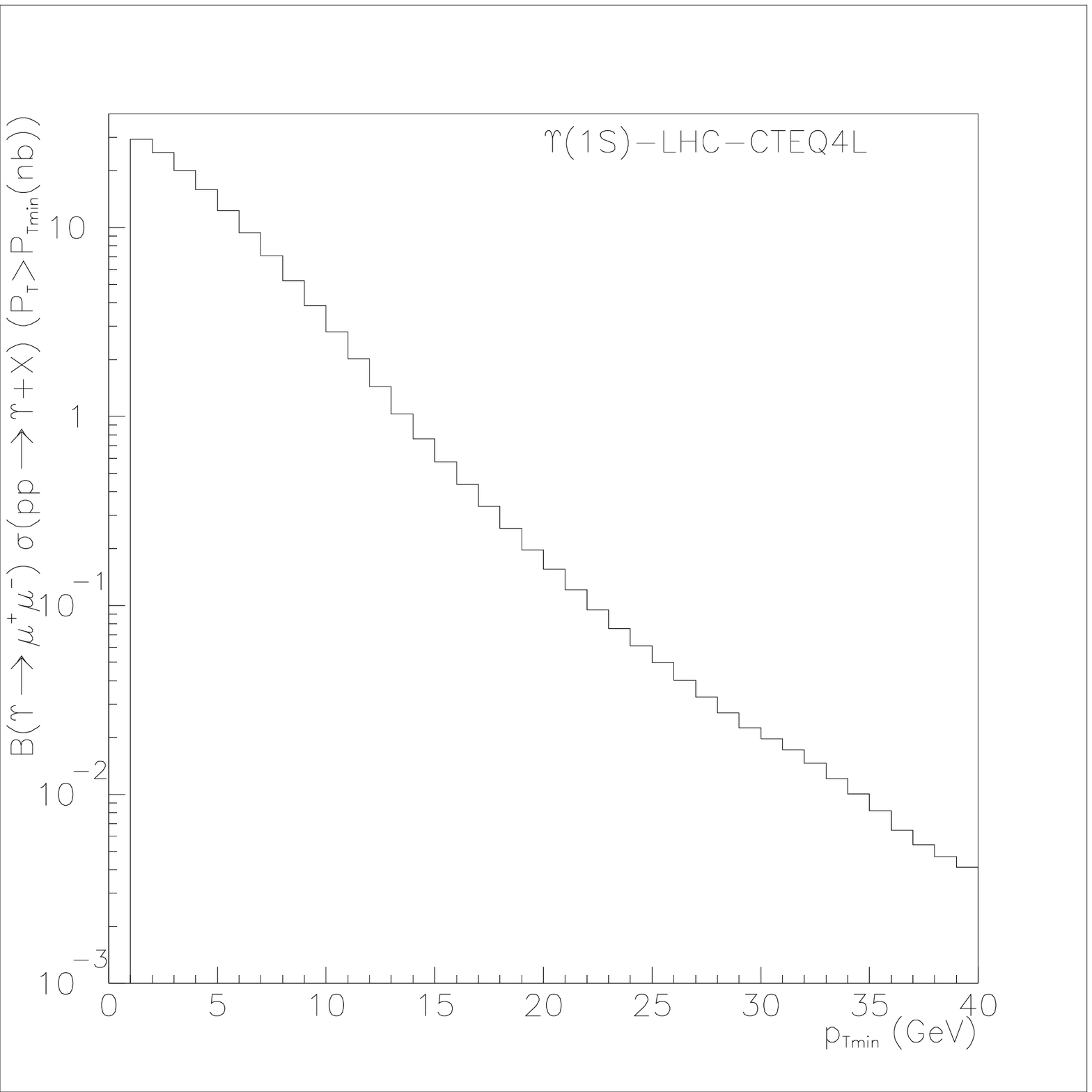,height=6.5cm,width=8.cm}
}}
\centerline{\hbox{
\psfig{figure=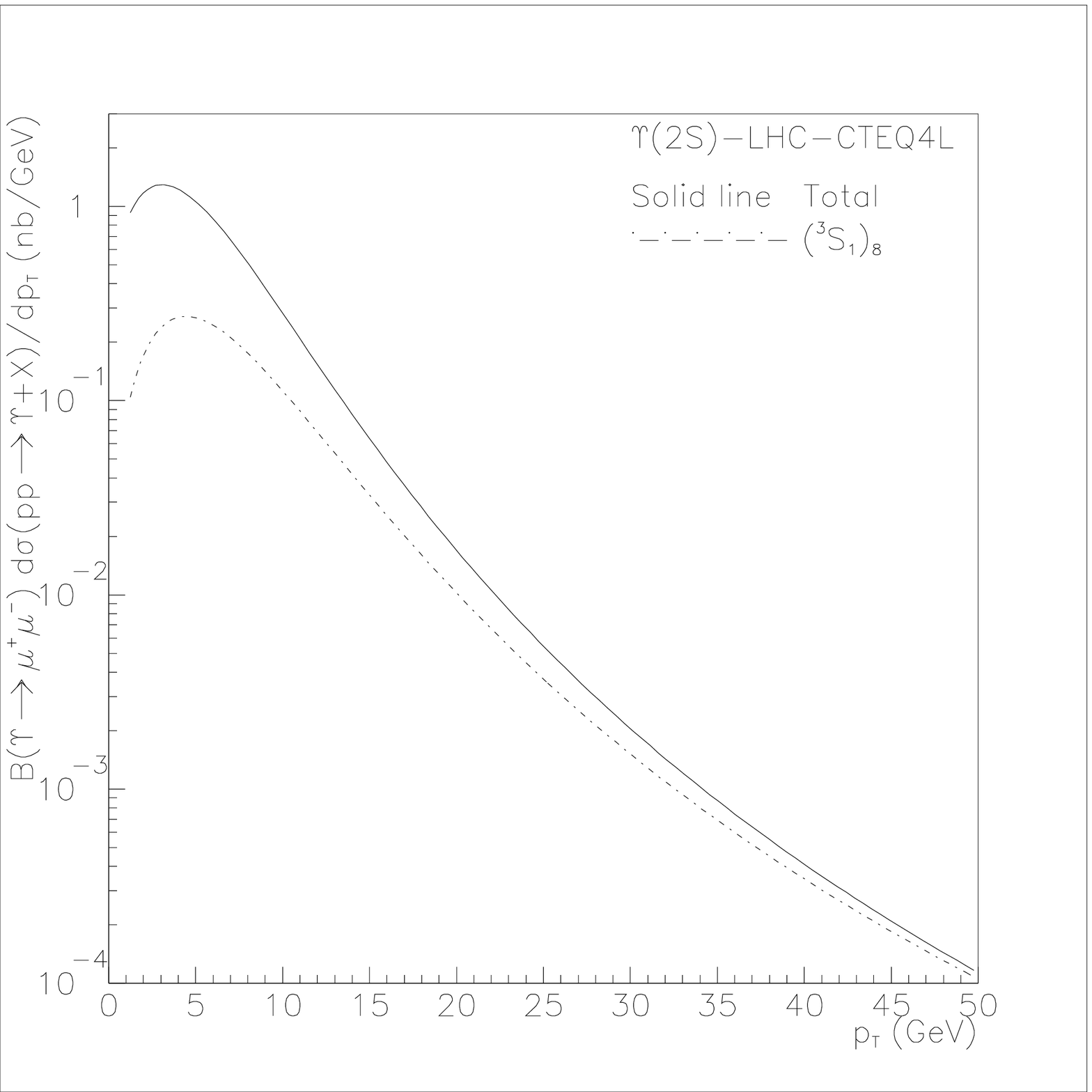,height=6.5cm,width=8.cm}
\psfig{figure=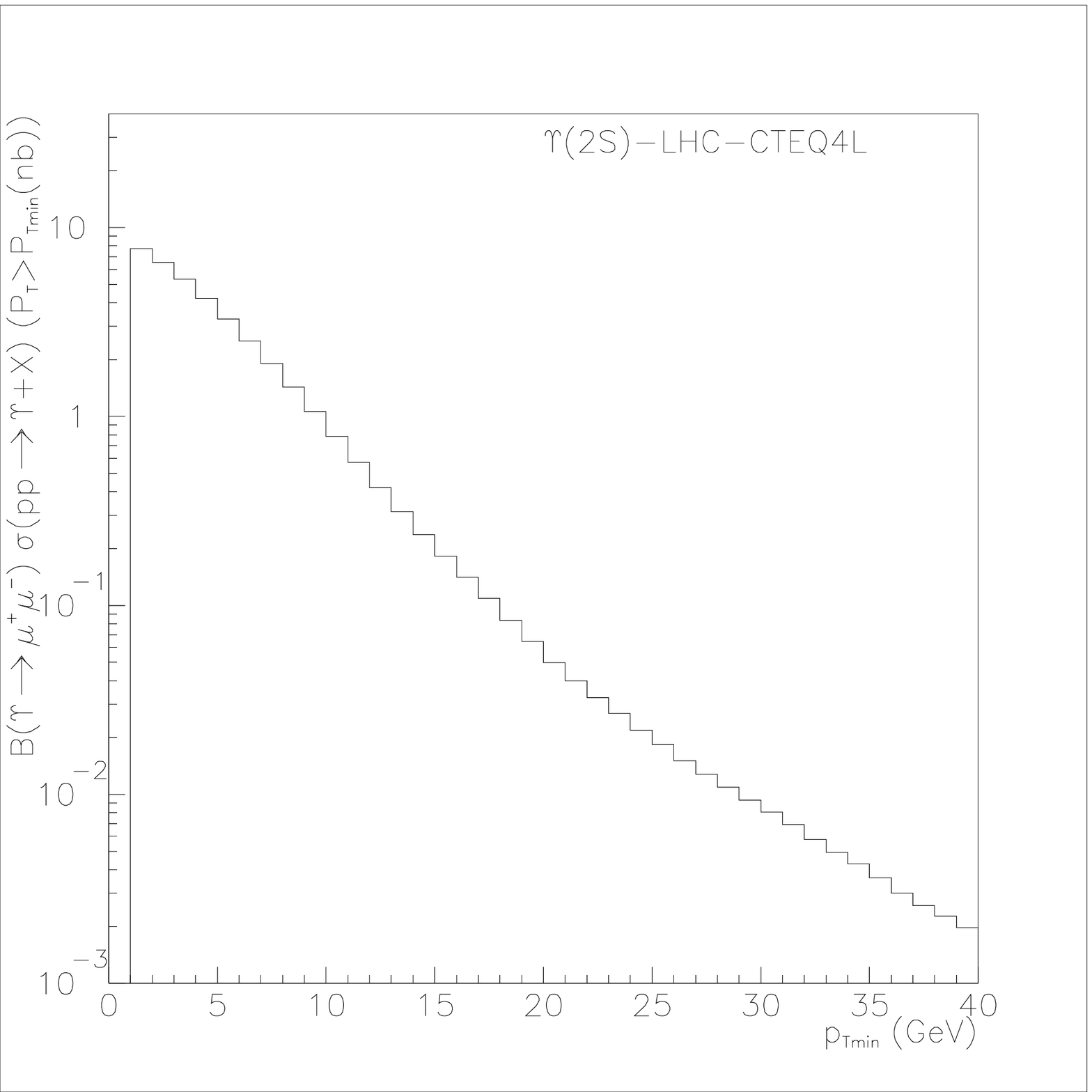,height=6.5cm,width=8.cm}
}}
\centerline{\hbox{
\psfig{figure=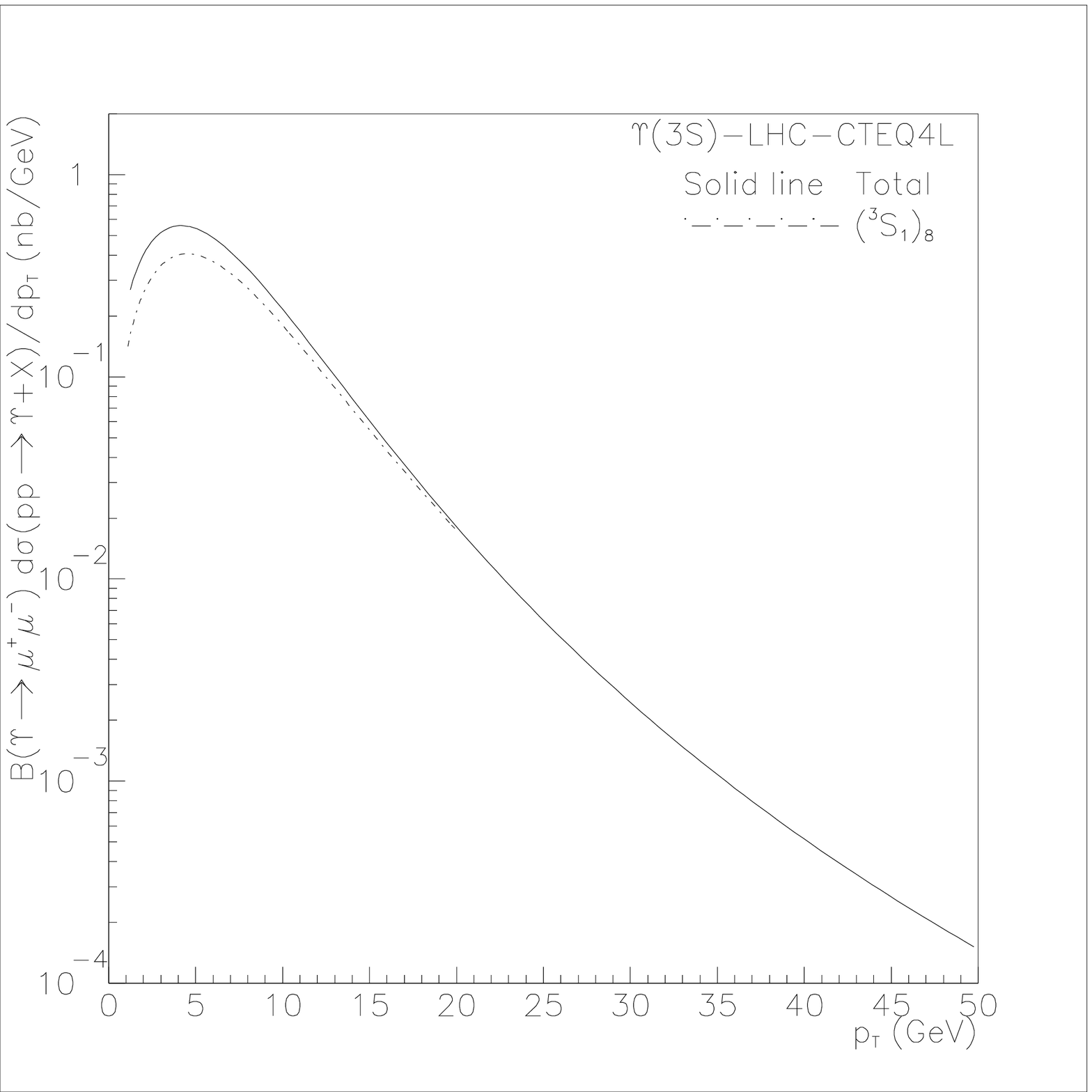,height=6.5cm,width=8.cm}
\psfig{figure=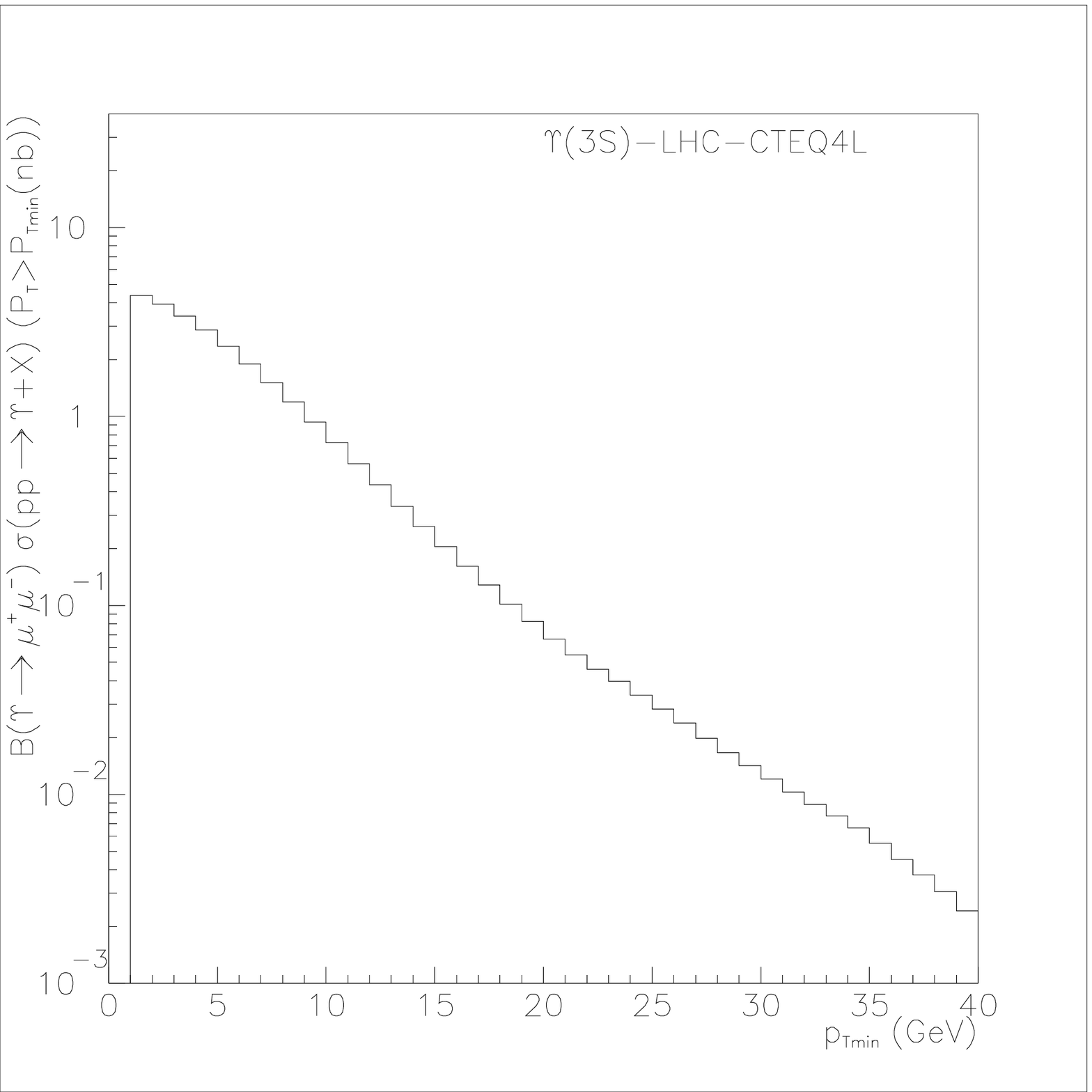,height=6.5cm,width=8.cm}
}}
\caption{{\em First row panels, left side} : Predicted prompt
$\Upsilon(1S)$ differential cross section (multiplied
by the muonic branching fraction) at the LHC
using the CTEQ4L PDF and $m_b=4.88$ GeV.
A rapidity cut ${\mid}y{\mid}<2.5$ was required for bottomonium. 
Dot-dashed line: $^3S_1^{(8)}$ contribution; solid line: all
contributions. {\em Right side :} Integrated cross section.
{\em Second and third row panels}: The same as in first row for 
$\Upsilon(2S)$ and $\Upsilon(3S)$ respectively.}
\end{figure}

\section{$\Upsilon(nS)$ Production at the LHC}

We already mentioned in the Introduction that bottomonium hadroproduction 
is especially  interesting to check the validity of the colour-octet model 
as often
emphasized in the literature \cite{beneke2,tkabladze}. This becomes 
particularly clear at the LHC since experimental data will spread over
a wider $p_T$-range than at the Tevatron, allowing an overall
study from low to very high $p_T$ values. Therefore
the expected transition of the different production mechanisms
along the $p_T$ region could be scrutinized in detail: from gluon gluon 
fusion at low $p_T$ to the foreseen asymptotically  
dominant gluon fragmentation into bottomonium states.

Keeping this interest in mind, we used our code implemented in
PYTHIA to generate prompt 
$\Upsilon(nS)$ resonances in proton-proton collisions at 
a center-of-mass energy of 14 TeV 
employing the best ${\chi}_{DF}^2$ colour-octet MEs shown in
Table 1. In figure 3 the theoretical curves for the 
$\Upsilon(nS)$ ($n=1,2,3$) differential and
integrated cross sections are exhibited as a function of $p_T$, including
both direct production and feeddown from higher resonance states
(except for the $\Upsilon(3S)$).

In Figures 4 we show our prediction for {\em direct} $\Upsilon(nS)$
production. This
is especially interesting if LHC detectors would be able to
discriminate among those different sources of resonance
production. (See the end of Section 4 and footnote $\#2$.)

To this end we generated $\Upsilon(1S)$ events
through both the CSM and COM making use of the following parameters
\begin{itemize}
\item $<O_1^{\Upsilon(1S)}(^3S_1)>{\mid}_{direct}=9.28$ GeV$^3$ 
(from \cite{schuler})
\item $<O_8^{\Upsilon(1S)}(^3S_1)>{\mid}_{direct}=0.084$ GeV$^3$
\end{itemize}

The first value corresponds to the CSM matrix element
for direct production while
the $<O_8^{\Upsilon(1S)}(^3S_1)>{\mid}_{dir}$ value was obtained after removing
the $\Upsilon(2S)+\Upsilon(3S)$ contribution according to
the discussion made in Section 2.1.1, i.e. under the assumption that
a fraction $7.6\%$ from the $36.8\%$ in table 3 should be assigned to 
indirect production. Finally let us mention that
we neglected any contribution from
the $^1S_0^{(8)}+^3P_J^{(8)}$ channels, in accordance with our analysis 
on Tevatron results of Section 2.

\vskip 1.cm
\begin{figure}[htb]
\centerline{\hbox{
\psfig{figure=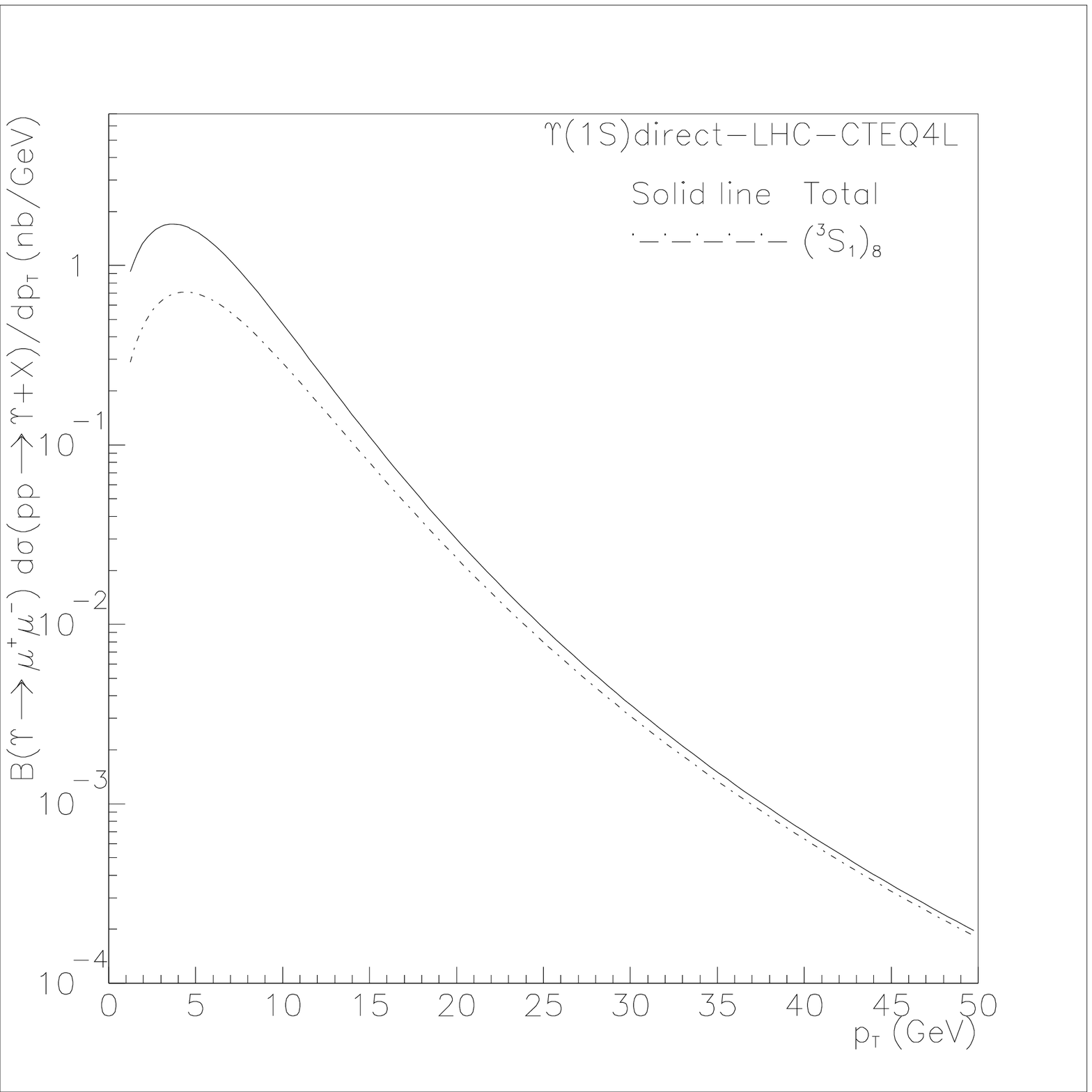,height=6.5cm,width=8.cm}
\psfig{figure=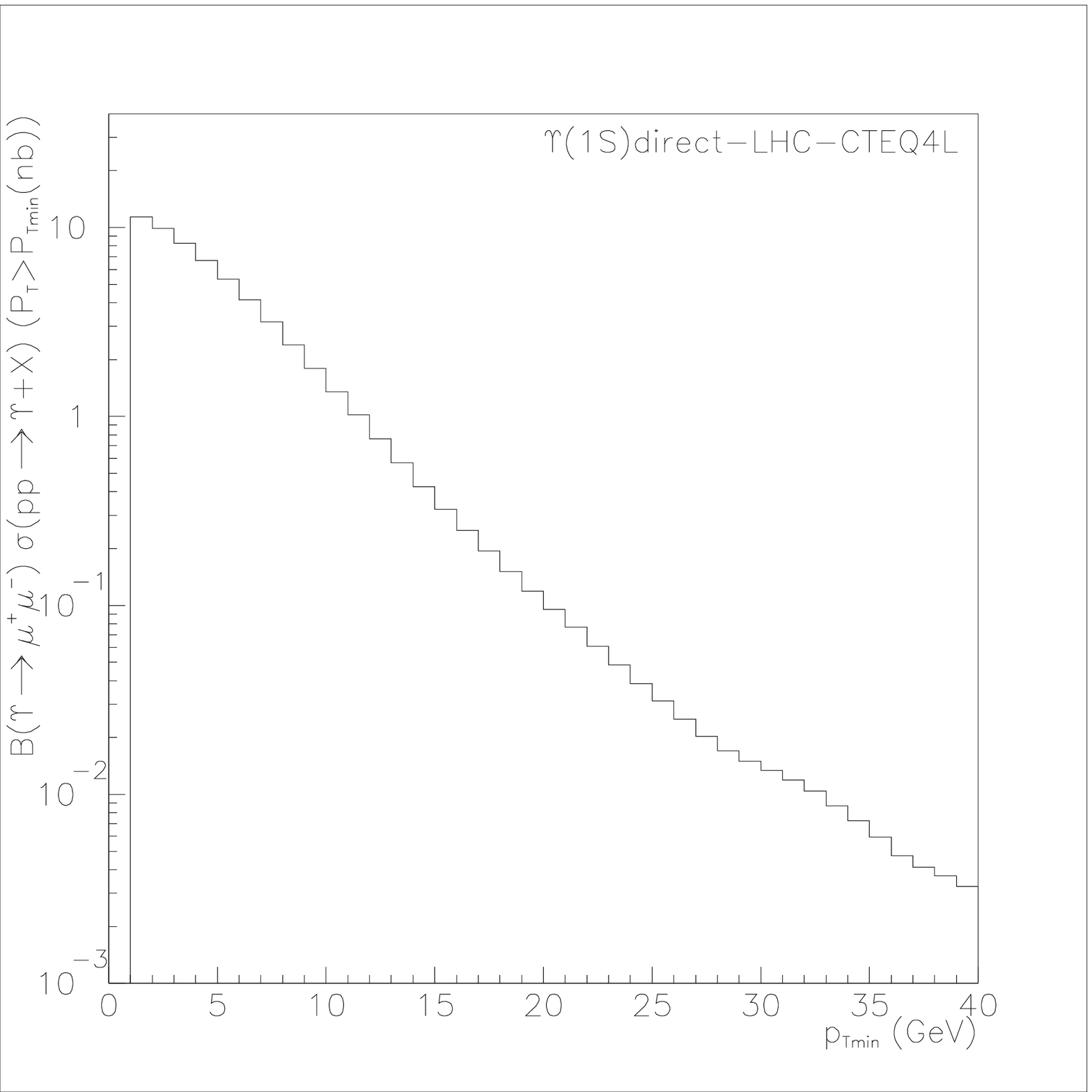,height=6.5cm,width=8.cm}
}}
\caption{The same as in Figure 3 for {\em direct} $\Upsilon(1S)$
production at the LHC.}
\end{figure}

\section{Heavy quarkonia inclusive production as a probe of the gluon density
in protons}

One of the goals of the LHC project is to perform precise tests of the
Standard Model of strong, weak and electromagnetic interactions and
the fundamental constituents of matter. In fact
the LHC machine can be viewed as a gluon-gluon collider to some extent.
On the other hand, 
many signatures (and their backgrounds) of physics, both within and 
beyond the Standard Model, involve gluons in the initial state
\cite{tdr}. Therefore an accurate 
knowledge of the gluon density in protons acquires a
special relevance for all these reasons. 

So far, the most precise determinations of the gluon momentum
distributions in the proton come from the analysis of the
scaling violations of the structure function $F_2$. However, 
this represents an indirect method since it is the sea distribution
which is actually measured and the gluon density is obtained
by means of the QCD evolution equations. On the other hand,  
hadron-hadron scattering processes with direct photon production
or jets in the final state will probably be extremely adequate to
probe $\lq\lq$directly'' the gluon distribution in hadrons.  
In this Section, we shall examine the possibility of using heavy quarkonia 
inclusive production in proton-proton collisions at the LHC, in a
complementary way to those studies. However, there are still
many uncertainties and pending questions regarding quarkonia 
hadroproduction despite the existence of NRQCD \cite{bodwin}, an 
effective theory
coming from first principles, which should provide an adequate
framework for this kind of processes involving both perturbative
and non-perturbative aspects of the strong interaction dynamics.
Likely, forthcoming experimental data - and their respective analyses - 
from Tevatron and other machines like HERA, should clarify the situation 
before LHC becomes operative.

In the following we shall focus on very high $p_T$ production
of bottomonia states. Therefore the main production mechanism 
according to the COM should be through the partonic subprocess:
\begin{equation}
g\ g \ {\rightarrow}\ g^{\ast}\ g
\end{equation}
followed by the gluon fragmentation into a $\Upsilon(nS)$ state:
\begin{equation}
 g^{\ast}{\rightarrow}\ \Upsilon(nS)\ X\ \ \ \ \ ;\ (n=1,2,3) 
\end{equation}
produced through a colour-octet mechanism. 
On the other hand, the bottom mass is large enough to justify
the colour-octet model applied to quarkonium hadroproduction. Whether
a similar approach could be applied to charmonium resonances has to
be checked, for example analyzing the transverse polarization of
the resonance. 

Ideally, the final state gluon ($g$) in Eq. (4)
will give rise to a recoiling jet ($g{\rightarrow}jet$), 
sharing, in principle, the same 
transverse momentum as the heavy resonance (in absence of higher order
corrections; see however appendix C). 
Hence events would topologically consist of an almost
isolated  muon pair from the decay
of the heavy resonance and a recoiling jet. Indeed one
should expect a ${\mu}^+{\mu}^-$ pair almost isolated because the
energy difference between the masses of the intermediate coloured and
final states is assumed to be rather small (of the order of 
$m_bv^2\ {\simeq}\  500$ MeV) then allowing the emission
of eventually a few light hadrons via soft gluon radiation at the
final hadronization stage.

Bottomonia production coming from fragmenting gluons in
QCD jets (an alternative production mechanism, see \cite{baranov2})
should not exactly display the same signature as the
hard ${\alpha}_s^3$ processes. Indeed, the muon  pair would be 
embedded in one of the two jets - not so much isolated as in
the process (4-5) due to the production cascade -  and its momentum 
should not balance
the momentum of the other event jet to the same extent.
In sum, the signature of an almost isolated muon pair recoiling against a jet
with an approximate momentum balance in the transverse plane, should provide
a suitable tag for the production mechanism represented in Eqs. (4-5).
 
We shall assume a tight kinematic cut in this our 
first approach: Both the rapidity of the heavy resonance and 
the rapidity associated to the
recoiling jet should be around zero. However, in order to
increase the foreseen statistics, one could dispense with this
constraint by only requiring (within the experimental and
theoretical uncertainties) back-to-back production. We
shall come back to this issue at the end of this Section.

\subsubsection*{Developing the idea}

In the absence of any intrinsic $k_T$ effect, we can write the 
triple differential cross 
section for the inclusive production process $pp{\rightarrow}{\Upsilon}X$ as
\begin{equation}
\frac{d^3\sigma}{dy_{\Upsilon}dy_{jet}dp_T}= 2p_T
\sum_{ab}x_ax_bf_{a/p}(x_a)f_{b/p}(x_b) \frac{d\hat{\sigma}_{ab}}{d\hat{t}}
\end{equation}
where $f_{a/p}(x_a)$ denotes the parton-$a$ density in the proton, and

\begin{equation}
\frac{d\hat{\sigma}_{ab}}{d\hat{t}}\ {\equiv}\ \frac{d\hat{\sigma}}{d\hat{t}}
(ab{\rightarrow}{\Upsilon}c)=\frac{1}{16{\pi}\hat{s}^2}\ 
\overline{\sum}{\mid}{\cal A}(ab{\rightarrow}{\Upsilon}c){\mid}^2
\end{equation}
stands for the partonic differential cross section (the barred summation
denotes an average over initial and final spins and colours)
consisting of a short distance (and calculable) part and a long distance 
part which can be identified as a colour-octet matrix element 
according to NRQCD. This factorization of the cross section was established 
on solid grounds in Ref. \cite{bodwin} within the NRQCD framework.

As above-mentioned we shall require 
both rapidities (of the $\Upsilon$ and the recoiling jet) to be 
less than a common small value $y_0$:
${\mid}y_{\Upsilon}{\mid}<y_0$, ${\mid}y_{jet}{\mid}<y_0$.
We could set $y_0=0.25$ for example, as discussed in appendix C.)
Then $x_a\ {\simeq}\ x_b=x$, and
\begin{equation}
x_ax_b=x^2=\frac{\hat{s}}{s}
\end{equation}
At very high $p_T$ (i.e. $p_T^2>>4m_b^2$) we can identify
$\hat{s}\ {\approx}\ 4p_T^2$. (Hereafter we consider $p_T\ {\ge}\ 20$ GeV.)
Therefore measuring the transverse momentum of the
resonance should lead to the the knowledge of the momentum fraction $x$ 
of the interacting partons, with a typical uncertainty (see appendix C)
\begin{equation}
\frac{{\Delta}x}{x}\ =\ y_0
\end{equation}

In particular the dominant partonic subprocess
should be the  gluon-gluon interaction. Thus the gluon density
$G(x,\mu^2)$ in the proton will mainly be involved and 
we can write as a first approximation

\begin{equation}
\frac{d^3\sigma}{dy_{\Upsilon}dy_{jet}dp_T}= 2p_T\ 
x^2\ G(x,\mu^2)^2\ \frac{d\hat{\sigma}_{gg}}{d\hat{t}}
\end{equation}
where we can choose, for example ${\mu}^2\ =\ \hat{s}$

\subsection*{The proposal
\footnote{Presented at the UK Phenomenology
Workshop on Heavy Flavour and CP violation, Durham, September 2000
\cite{mas6}
and at the $B$ physics working group meeting of the ATLAS 
collaboration held at CERN in October, 2000 \cite{mas7}.}}

We propose to study the ratios:
\begin{equation}
\frac{x_2^2\ G(x_2,\mu_2^2)^2}{x_1^2\ G(x_1,\mu_1^2)^2}\ =\  
\biggl(\frac{d\hat{\sigma}_{gg}/d\hat{t}_1}{d\hat{\sigma}_{gg}/d\hat{t}_2}\biggr)
{\times}
\biggl(\frac{p_{T1}}{p_{T2}}\biggr){\times}
\biggl(\frac{d^3{\sigma}/dy_{\Upsilon}dy_{jet}dp_{T2}}
{d^3{\sigma}/dy_{\Upsilon}dy_{jet}dp_{T1}}\biggr)
\end{equation}
for a set of $x_1,x_2$ pairs and
{\em different gluon distributions}. 
The number of pairs is basically limited by ${\Delta}x$, i.e. $y_0$, so
this constraint cannot released too much (see appendix C).

Therefore the keypoint is to consider the l.h.s. of the above equality
(Eq. (11)) as an {\em input} corresponding to different sets of the
gluon distribution for the proton, whose $x$ dependence is hence
assumed to be $\lq\lq$known'', and in fact would be tested.
On the other hand the r.h.s. corresponds to an input from experimental
data and some theoretical factors likely under control.

Let us remark
that the $x$ and ${\mu}^2$ values are {\em not independent}
in this proposal; indeed for each value of $x$, ${\mu}^2$
is fixed by $\hat{s}=x^2s$. However, notice that the
scale can actually be varied by choosing a different assignment
for ${\mu}^2$, e.g. ${\mu}^2=\hat{s}/4$.

Next we shall write expression (11) as
\begin{equation}
\frac{x_2^2\ G(x_2,\mu_2^2)^2}{x_1^2\ G(x_1,\mu_1^2)^2}\ =\ 
R_{theo}\ {\times}\ R_{exp}
\end{equation}
where
\begin{equation}
R_{theo}(p_{t1},p_{t2},{\mu}_1^2,{\mu}_2^2)\ =\ f_{cor}{\times}
\frac{d\hat{\sigma}_{gg}/d\hat{t}_1}{d\hat{\sigma}_{gg}/d\hat{t}_2}
\end{equation}
and in the high $p_T$ limit,
\[ R_{theo}\ {\rightarrow}\ 
f_{cor}\ {\times}\ \frac{{\alpha}_s^3(\mu_1^2)\ p_{T2}^4}
{{\alpha}_s^3(\mu_2^2)\ p_{T1}^4}  \]
explicitly showing that ${\alpha}_s(\mu^2)$ is entangled in the
gluon density determination. On the other hand, 
note that the dependence on the NRQCD matrix elements does cancel
in $R_{theo}$, but there is a dependence on the scales
$\mu_1^2$ and $\mu_2^2$, which should match the same dependence
in the left hand side.
We have incorporated some possible corrections through the
$f_{cor}$ factor - which could be calculated either analytically
or by Monte Carlo methods -  taking into
account higher-order effects such as intrinsic $k_T$ of 
the interacting gluons,
AP evolution of the fragmenting gluons, etc.

On the other hand the experimental input reads as the ratio

\begin{equation}
R_{exp}(p_{T1},p_{T2},y_0)= 
\biggl(\frac{p_{T1}}{p_{T2}}\biggr)\ {\times}\ 
\biggl(\frac{d^3{\sigma}/dy_{\Upsilon}dy_{jet}dp_{T2}}
{d^3{\sigma}/dy_{\Upsilon}dy_{jet}dp_{T1}}\biggr)
\end{equation}
which can be obtained directly from experimental data.

\subsubsection*{Introducing the gluon quark contribution}

Although expectedly dominant, the gluon gluon partonic subprocess 
is not the only $\alpha_s^3$ contribution to the  
cross section yielding a fragmenting gluon into $\Upsilon(nS)$ 
at high $p_T$. Also gluon quark scattering
$gq{\rightarrow}g^{\ast}q$ followed by
$g^{\ast}{\rightarrow}\Upsilon(nS)X$, can give a sizeable contribution
(about $20\%$ at $p_T>20$ GeV, see table 6 in appendix A). Consequently, 
the expression (12) for the ratio of gluon densities
has to be modified to include the quark distribution 
$q(x,\mu^2)$ in protons:

\begin{equation}
\frac{x_2G(x_2,\mu_2^2)\  
(x_2G(x_2,\mu_2^2)+k{\cdot}x_2q(x_2,\mu_2^2))}
{x_1G(x_1,\mu_1^2)\ (x_1G(x_1,\mu_1^2)+k{\cdot}x_1q(x_1,\mu_1^2))}\ =\  
R_{theo}\ {\times}\ R_{exp}
\end{equation}
where $k$ is a factor taking into account the 
ratio of the $gq$ and $gg$ cross sections, both calculated at the same 
values of the Mandelstam variables $\hat{s}$ and $\hat{t}$ of the
hard interaction, i.e.

\begin{equation}
k\ =\ \frac{d\hat{\sigma}_{gq}/d\hat{t}}{d\hat{\sigma}_{gg}/d\hat{t}} 
\end{equation}
becoming independent of $x$ (and $\mu^2$) at zero rapidity
and large $p_T$; then $k\ {\simeq}\ 0.2$.

Alternatively, one can write the density ratio as

\begin{equation}
\frac{x_2^2\ G(x_2,\mu_2^2)^2 
(1+k{\cdot}\lambda(x_2,\mu_2^2))}
{x_1^2\ G(x_1,\mu_1^2)^2
(1+k{\cdot}\lambda(x_1,\mu_1^2))}
\end{equation}
where
\[
\lambda(x,\mu^2)=\frac{q(x,\mu^2)}{G(x,\mu^2)} 
\]

By Taylor expanding the above ratio, the leading term is

\[
\frac{x_2^2\ G(x_2,\mu_2^2)^2}{x_1^2\ G(x_1,\mu_1^2)^2}\ 
(1+r) \]
where

\[ r=k{\times}
\biggl[\lambda(x_2,\mu_2^2)-\lambda(x_1,\mu_1^2)\biggr]
\]
should be a quite small quantity. (We have checked with 
CTEQ4L that typically $r\ {\approx}\ 0.03$
for values between $x_1=3{\cdot}10^{-3}$ and $x_2=1.5{\cdot}10^{-2}$.)

We can rewrite Eq. (12) as

\begin{equation}
\frac{x_2^2\ G(x_2,\mu_2^2)^2}{x_1^2\ G(x_1,\mu_1^2)^2}
(1+r)=R_{theo}\ {\times}\ R_{exp}
\end{equation}
Again the l.h.s. is an input from the PDF to be tested, 
while the r.h.s. comes from experimental data and some 
theoretical calculations
without requiring the NRQCD MEs values.

From an experimental point of view it may happen that 
the discrimination among the different 
$\Upsilon(nS)$ states via mass reconstruction could become 
a difficult task, especially at very high $p_T$, 
because of the uncertainty on the measurement of the muons momenta \cite{tdr}
\footnote{We thank M. Smizanska for bringing this
point to our attention.}. Nevertheless, since we are
proposing to study {\em ratios} of cross sections, we can consider 
the overall $\Upsilon(nS)$ inclusive production, without separating 
the different bottomonia sources - all the weighted matrix element cancelling
in the quotient if we neglect the mass differences
between the different states. (Notice that at high $p_T$ there is almost
no contribution from the CSM.) In Figure 5 we show the combined
production rate at $p_T>20$ GeV and ${\mid}y{\mid}<0.25$
for the upper and lower values of the colour-octet
matrix elements shown in Table 1.

\begin{figure}[htb]
\centerline{\hbox{
\psfig{figure=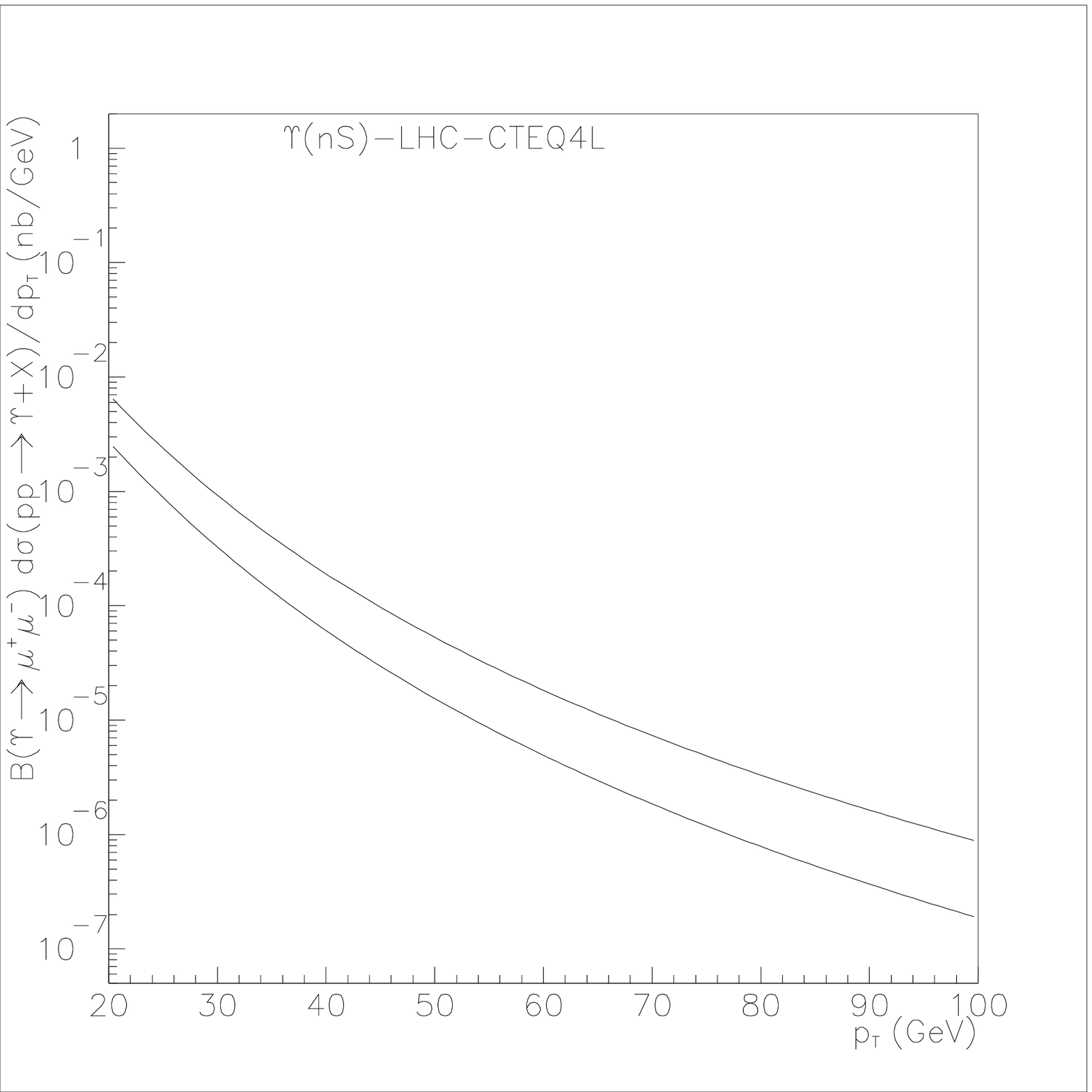,height=6.5cm,width=8.cm}
\psfig{figure=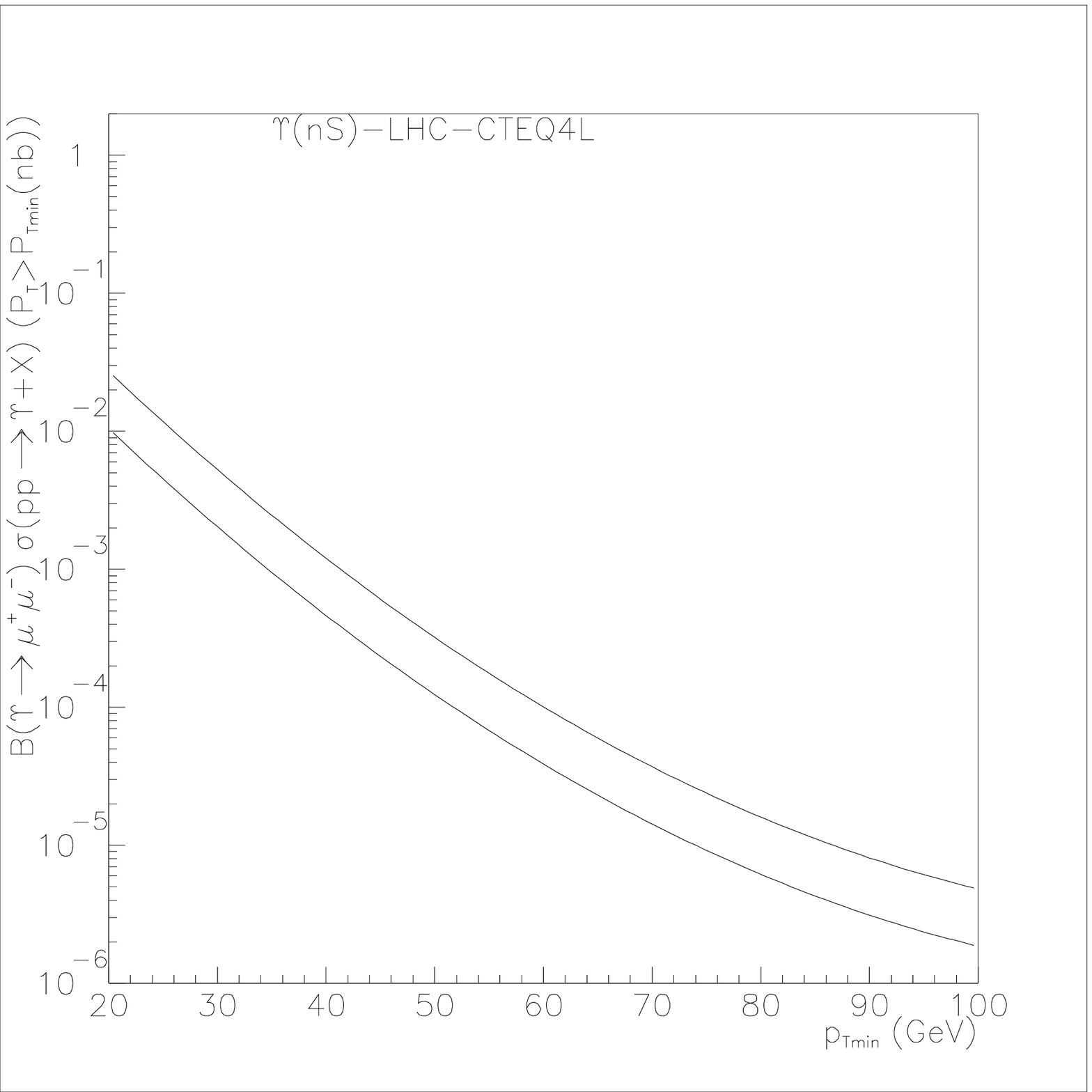,height=6.5cm,width=8.cm}
}}
\caption{Predicted  $\Upsilon(1S)+\Upsilon(2S)+\Upsilon(3S)$ weighted
contributions to bottomonia inclusive production at the LHC 
corresponding to the upper and lower MEs from Table 1, in the
rapidity interval $|y|<0.25$ and $p_T>20$ GeV.
{\em Left panel}: differential cross section; {\em right panel}: integrated
cross section.}
\end{figure}

Assuming an integrated luminosity of $10$ fb$^{-1}$, corresponding
to one year running ($10^{7}$s) of LHC at $\lq\lq$low'' luminosity
($10^{33}$ cm$^{-2}$s$^{-1}$) we can easily get the expected number
of events from figures 5, just by multiplying the ordinate by
a factor $10^7$. Thus we can see that the foreseen number
of events (aside efficiency reduction) at $p_T>20$ GeV
is about $10^{5}$, whereas at  $p_T>40$ GeV
is about $10^{4}$. By extrapolation  we get a meagre expected number of  
${\simeq}\ 10^{2}$ events at $p_T>100$ GeV. This makes unlikely any
measurement for transverse momentum larger than 100 GeV, under
the tight rapidity cut of $0.25$ on the resonance which we are imposing.

In view of the foreseen rates of bottomonia production at the LHC we 
propose testing the shape of the gluon density in protons for
$x$ values ranging in the 
interval: $3{\times}10^{-3}\ to\  1.5{\times}10^{-2}$, 
using $x=\sqrt{\hat{s}/s}$ from $p_T=20$ GeV up to $p_T=100$ GeV, 
under the rapidity constraint $y<0.25$.

Nevertheless, by removing the condition ${\mid}y{\mid}<0.25$ statistics 
could considerably be enlarged. Since our proposal
essentially relies on the determination of the Feynman $x$
of the interacting partons by measuring the $p_T$ of the
final products of the reaction, there is still the possibility
of requiring a back-to-back topology but sweeping
the whole
accessible rapidity region ${\mid}y{\mid}<2.5$, instead of limiting 
ourselves to the central
rapidity values. This goal can be achieved by
selecting events with the muon pair and the recoiling
jet sharing common values of $p_T$ and absolute
rapidities, within the uncertainties. In other words, events could be accepted
with both $\Upsilon(nS)$ and recoiling jet rapidities 
satisfying ${\mid}{\mid}y_{\Upsilon}{\mid}-{\mid}y_{jet}{\mid}{\mid}
<0.25$;
in such a way statistics should increase by a factor 
${\simeq}\ 10$, possibly extending the allowed region of $p_T$
up to higher values than 100 GeV, and hence 
reaching larger values of the momentum fraction $x$.

As a final remark, if the colour-octet model is confirmed and the
corresponding MEs accurately and consistently  
extracted from other experiments like Tevatron
or HERA - or theoretically computed - one can consider then
the possibility of unfolding the gluon density from the measured
cross section as proposed, for instance, in Ref. \cite{H1} by
means of $D^{\ast}$ meson production at HERA. In such a case,
our proposal would extend 
beyond the study of ratios, allowing the extraction of
gluon and quark densities directly from heavy quarkonia production
mechanisms.

\section{Summary}

In this paper we have analyzed CDF measurements on $\Upsilon(nS)$
($n=1,2,3$) inclusive hadroproduction cross sections at the Tevatron in a 
Monte Carlo framework, extracting some relevant
colour-octet NRQCD matrix elements. Higher-order QCD effects such as 
initial-state radiation and AP evolution of gluons were included
in our analysis. The fact that we were able to take into account
the effects of soft-gluon emission in the initial-state (according
to the PYTHIA machinery) yielding a smooth $p_T$ behaviour
at low $p_T$, allowed us to extend the study of the differential
cross section down to $p_T=2$ GeV.
However, in order to study the sensitivity to different cut-offs, 
fits were performed using experimental points above
several $p_T$ cut-offs (see Table 1).

On the other hand, since the different sources 
of $\Upsilon(nS)$ production were not 
experimentally separated along the full accessible $p_T$-range
we have included all of them in the generation and later fits. Only 
for $p_T>8$ GeV, feeddown from $\chi_{bJ}$ states was 
experimentally separated out from
direct $\Upsilon(1S)$ production. We used these results as a 
consistency check of our analysis and to draw some conclusions
summarized below.

The numerical value of the
$<O_8^{\Upsilon(1S)}(^3S_1)>{\mid}_{tot}$ matrix element
should be ascribed almost totally to ${\Upsilon(nS)}$ states. 
This finding may be surprising when confronted with other
analyses \cite{cho,schuler}, 
where the contribution to the ${\Upsilon(1S)}$ yield
through the colour-octet
$\chi_{bJ}$ channels was thought as dominant 
\cite{schuler,tkabladze,beneke}. 
On the contrary, we concluded from Tables 2
and 3 that the {\em colour-singlet production} can account 
by itself for the feeddown of $\Upsilon(1S)$ from  ${\chi}_{bJ}$
states. (Notice however that experimental uncertainties still
leave some room for a possible COM contribution but to a much lesser
extent than previously foreseen \cite{cho,schuler}.)
On the other hand the different production channels
are consistent (or can be made consistent)
with the experimental relative fractions shown in Table 2, 
after some reasonable assumptions.

We have extended our study to LHC collider
experiments ($\sqrt{s}=14$ TeV center-of-mass energy).
In Figure 3  we presented our predictions for prompt
production rates (i.e. including direct and indirect
production) while in Figure 4 we showed our prediction
for direct $\Upsilon(1S)$ production alone.

We conclude that the foreseen yield of $\Upsilon(nS)$'s
at LHC energy will be large enough, even at
high-$p_T$, to perform a detailed analysis of the colour-octet
production mechanism and should be 
included in the B-physics programme of the LHC experiments, 
probably deserving (together with charmonia) a dedicated
data-taking trigger.

On the other hand, we have presented in some detail 
(but without considering detector effects) the prospects
to use heavy quarkonia inclusive production at the LHC 
(or perhaps at the Tevatron too, using data from the
high luminosity Run II) to probe the
gluon density in protons.
Let us remark that we are renouncing to use any absolute normalization 
because this would imply a precise knowledge of 
the colour-octet NRQCD
matrix elements governing the transition of the fragmenting gluon into
a particular heavy quarkonium state, hence introducing an important
uncertainty since they are not accurately known so far. 
If those NRQCD elements were finally accurately and consistently
determined (either theoretically or experimentally) one could then
consider the possibility of unfolding the parton densities from
LHC experimental data on heavy resonance inclusive production, as 
described in our proposal. Thus we conclude that: {\em a)}
Inclusive hadroproduction of heavy resonances at high $p_T$ in
general-purpose LHC experiments
could be a complementary method of constraining the gluon density in
protons, along with other related
methods: di-jet, lepton pair and prompt photon production;
{\em b)} An experimental advantage of this method lies in the fact that 
the $\lq\lq$flight direction'' of a high-$p_T$ fragmenting gluon
into a $\Upsilon(nS)$ resonance can be inferred
from the muonic pair coming from its decay, thereby providing a 
clean constraint in the search
for the associated recoiling jet, in addition to the
self-triggering signature of events.

We finally want to stress the importance of keeping an open mind on
the different possibilities offered by the LHC, 
likely deserving a previous work to prepare in advance
jointly experimental strategies and theoretical calculations. 
\vskip 0.8cm

\subsection*{Acknowledgements}
We acknowledge interesting discussions with N. Ellis,  
S. Frixione, M. Kr\"{a}mer, F. Maltoni, M. Smizanska,
Y. Tsipolitis and S. Wolf.

\newpage

\appendix\renewcommand{\theequation}{\thesection.\arabic{equation}}
\section*{Appendices} 

\appendix{}

\section{}

\setcounter{equation}{0}

\large{\bf Some technical details for the generation with PYTHIA}

Basically we reproduce here the information given in \cite{mas4} about the
values of the parameters and options set in running PYTHIA, with some
more details.

Originally the event generator PYTHIA 5.7 produces direct $J/\psi$ and
higher ${\chi}_{cJ}$ resonances via the CSM only \cite{pythia}. It is
not a difficult task to extend this generation to the bottomonium family
by redefining the resonance mass and wave function parameter accordingly.
In our analysis we have besides implemented a code 
in the  event  generator to account for
the colour-octet production mechanism via the following
${\alpha}_s^3$ partonic processes:

\begin{equation}
g\ +\ g\ {\rightarrow}\ (Q\overline{Q})[^{2S+1}X_J]\ +\ g
\end{equation}
\begin{equation}
g\ +\ q\ {\rightarrow}\ (Q\overline{Q})[^{2S+1}X_J]\ +\ q
\end{equation}
\begin{equation}
q\ +\ \overline{q}\ {\rightarrow}\ (Q\overline{Q})[^{2S+1}X_J]\ +\ g
\end{equation}
\vskip 0.05cm
where $(Q\overline{Q})[^{2S+1}X_J]$ stands for a certain heavy
quarkonium state denoted by its spectroscopic notation
(see Refs. \cite{cho,mas2} for more details). In particular
we have considered  the
$^3S_1^{(8)}$, $^1S_0^{(8)}$ and $^3P_J^{(8)}$ contributions
as leading-order intermediate coloured states. In addition we generated
$\Upsilon(nS)$ $(n=1,2,3)$ and $\chi_{bJ}(nP)$ ($n=1,2$) resonances 
decaying into $\Upsilon(1S)$, according to the CSM as mentioned above. 
\vskip .8cm
\begin{figure}[htb]
\centerline{
\psfig{figure=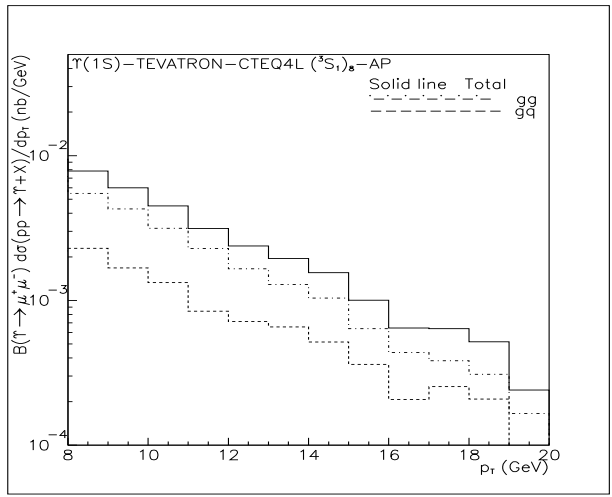,height=6.5cm,width=8.cm}
}
\caption{Gluon-gluon versus quark-gluon $^3S_1^{(8)}$ contributions from our
$\Upsilon(1S)$ generation at the Tevatron for $p_T>8$ GeV. The
latter becomes more and more important at larger $p_T$ as could be expected
since higher Feynman $x$ of partons in the proton are involved and the
$gq$ contribution becomes increasingly more significant w.r.t. 
the $gg$ one.}
\end{figure}

\begin{table*} [htb]
\setlength{\tabcolsep}{1.5pc}
\caption{$^3S_1^{(8)}$ contributions to the $\Upsilon(1S)$  cross section at 
the Tevatron for $p_T>8$ GeV}
\begin{center}
\begin{tabular}{cc}
\hline
 Contribution       & $ \%$       \\
\hline
 $ gg  $       & $69$        \\
\hline
 $ qg  $      &  $30$        \\
\hline
 $ q\overline{q}  $       & $1 $        \\
\hline
\end{tabular}
\end{center}
\end{table*}

\begin{table*} [htb]
\setlength{\tabcolsep}{1.5pc}
\caption{$^3S_1^{(8)}$ contributions to the $\Upsilon(1S)$ cross section at 
the LHC for $p_T>8$ GeV}
\begin{center}
\begin{tabular}{cc}
\hline
Contribution       & $ \%$       \\
\hline
 $ gg  $       & $80$        \\
\hline
 $ qg  $      &  $20$        \\
\hline
 $ q\overline{q}  $       & ${\simeq}\ 0 $        \\
\hline
\end{tabular}
\end{center}
\end{table*}

A $p_T$ lower cut-off was set equal to 1 GeV (by default in PYTHIA)
throughout  the generation
since some of the contributing channels are singular at vanishing
transverse momentum \cite{montp}. Furthermore, all fits of Tevatron
data were performed using $p_T$ values above 2 GeV. 

We find from our simulation (see Tables 5 and 6) 
that gluon-gluon scattering actually stands 
for the dominant process at high $p_T$ as expected, gluon-quark scattering 
contributes appreciably however (${\simeq}\ 20-30\%$ of the colour-octet
production cross section) whereas the
quark-antiquark scattering represents a quite small
fraction ( ${\simeq}\ 1\%$ at the
Tevatron). In Figure 6 we plot the gluon-gluon and quark-gluon $^3S_1^{(8)}$
contributions as a function of the transverse momentum 
of the resonance obtained from our generation for the Tevatron.
This kind of information could be particularly interesting
for our discussion on the probe of the gluon density in protons
developed in Section 4. 
\newline
\vskip 0.5cm

\subsection*{Set of $\lq\lq$fixed'' and free parameters used in the generation}

Below we list the main parameters, including 
masses and branching fractions,  used in our generation with PYTHIA 5.7.
We employed the CTEQ4L parton distribution function (PDF) in all our 
\vspace{0.2in} analysis.
\newline
{\em Masses and branching fractions:}
\begin{itemize}
\item $m_b=4.88$ GeV
\item $m_{resonance}=2m_b$ in the theoretical calculation of the 
cross sections for the short-distance processes \cite{cho}. In the phase 
space factors for the event
generation we set, however, real masses of the $\Upsilon(nS)$
and weighted mean values for the ${\chi}_b(nP)$ resonances \cite{mas4}. 
\item $BR[\Upsilon(1S){\rightarrow}\mu^+\mu^-]=2.48\ \%$ (\cite{pdg})
\item $BR[\Upsilon(2S){\rightarrow}\mu^+\mu^-]=1.31\ \%$ (\cite{pdg})
\item $BR[\Upsilon(3S){\rightarrow}\mu^+\mu^-]=1.81\ \%$ (\cite{pdg})
\end{itemize}

\vskip 0.5 cm
{\em Colour-singlet parameters} (from \cite{schuler}):
\begin{itemize}
\item $<O_1^{\Upsilon(1S)}(^3S_1)>{\mid}_{tot}=11.1$ GeV$^3$
\item $<O_1^{\Upsilon(2S)}(^3S_1)>{\mid}_{tot}=5.01$ GeV$^3$
\item $<O_1^{\Upsilon(3S)}(^3S_1)>{\mid}_{tot}=3.54$ GeV$^3$, defined as
\begin{equation}
<O_1^{\Upsilon(nS)}(^3S_1)>{\mid}_{tot}\ =\ \sum_{m \geq n}^3
<O_1^{\Upsilon(mS)}(^3S_1)>Br[\Upsilon(mS){\rightarrow}\Upsilon(nS)X]
\end{equation}
\item $<O_1^{\chi_{b1(1P)}}(^3P_1)>=6.09$ GeV$^5$
\item $<O_1^{\chi_{b1(2P)}}(^3P_1)>=7.10$ GeV$^5$
\end{itemize}
\vskip 0.5cm
The radial wave functions at the origin (and their derivatives) 
used in the generation can
be related to the above matrix elements as
\begin{equation}
<O_1^{\Upsilon(nS)}(^3S_1)>\ =\ \frac{9}{2\pi}{\mid}R_{n}(0){\mid}^2
\end{equation}
\begin{equation}
<O_1^{\chi_{bJ(nP)}}(^3P_J)>\ =\ \frac{9}{2\pi}(2J+1) {\mid}R_n'(0){\mid}^2
\end{equation}
whose numerical values were obtained from
a Buchm\"{u}ller-Tye potential model tabulated in Ref. 
\vspace{0.2in} \cite{eichten}. \newline
{\em Colour-octet long-distance parameters to be extracted from the fit:}
\newline
\begin{itemize}
\item $<O_8^{\Upsilon(nS)}(^3S_1)>{\mid}_{tot}$, defined as
\begin{eqnarray}
<O_8^{\Upsilon(nS)}(^3S_1)>{\mid}_{tot} & & =\ \sum_{m \geq n}^3
<O_8^{\Upsilon(mS)}(^3S_1)>Br[\Upsilon(mS){\rightarrow}\Upsilon(nS)X]
\nonumber \\ 
& & +\ \sum_{m \geq n }^2\ \sum_{J=0}^2
<O_8^{\chi_{bJ}(mP)}(^3S_1)>Br[\chi_{bJ}(mP){\rightarrow}\Upsilon(nS)X] \nonumber \\
\end{eqnarray}
\item $<O_8^{\Upsilon(nS)}(^1S_0)>{\mid}_{tot}$, defined as
\begin{eqnarray}
<O_8^{\Upsilon(nS)}(^1S_0)>{\mid}_{tot} & & =\ \sum_{m \geq n}^3
<O_8^{\Upsilon(mS)}(^1S_0)>Br[\Upsilon(mS){\rightarrow}\Upsilon(nS)X]
\nonumber \\ 
\end{eqnarray}

\item $<O_8^{\Upsilon(nS)}(^3P_0)>{\mid}_{tot}$, defined as
\begin{eqnarray}
<O_8^{\Upsilon(nS)}(^3P_0)>{\mid}_{tot} & & =\ \sum_{m \geq n}^3
<O_8^{\Upsilon(mS)}(^3P_0)>Br[\Upsilon(mS){\rightarrow}\Upsilon(nS)X]
\nonumber \\ 
\end{eqnarray}
\end{itemize}
\vskip 0.5cm
\par
On the other hand, the differences in shape between the
$^1S_0^{(8)}$ and $^3P_J^{(8)}$ contributions were not sufficiently great
to justify independent generations for them. In fact, 
temporarily setting $<O_8^{\Upsilon(1S)}(^3P_0)>
=m_b^2<O_8^{\Upsilon(1S)}(^1S_0)>$ and
defining the ratio 
\begin{equation}
r(p_T)\ =\ \frac{{\sum}_{J=0}^{2}\frac{d{\sigma}}{dp_T}[^3P_J^{(8)}]}
{\frac{d{\sigma}}{dp_T}[^1S_0^{(8)}]}
\end{equation}
it is found $r\ {\simeq}\ 5$ as a mean value over the $[0,20]$ GeV $p_T$-range.
Actually the above ratio is not steady as a function of
the $\Upsilon(1S)$ transverse momentum. Therefore in the generation we 
split the $p_T$ region into two domains:
for $p_T\ {\leq}\ 6$ GeV we set  $r= 6$ whereas for $p_T>6$ GeV we set
$r=4$. 

In summary, only the $^1S_0^{(8)}$ channel was generated but
rescaled by the factor $r$ to incorporate the $^3P_J^{(8)}$
contribution as we did in \cite{mas2} for charmonium hadroproduction.
Consequently, in analogy to \cite{cho} 
we shall consider only the 
combination of the colour-octet matrix elements: 

\begin{equation}
M_5\ =\ 5\ {\times}\ \biggl(\frac{<O_8^{\Upsilon(1S)}(^1S_0)>{\mid}_{tot}}{5}+
\frac{<O_8^{\Upsilon(1S)}(^3P_0)>{\mid}_{tot}}{m_b^2}\biggr)
\end{equation}

\newpage

\vskip 0.5cm
\begin{figure}[htb]
\centerline{\hbox{
\psfig{figure=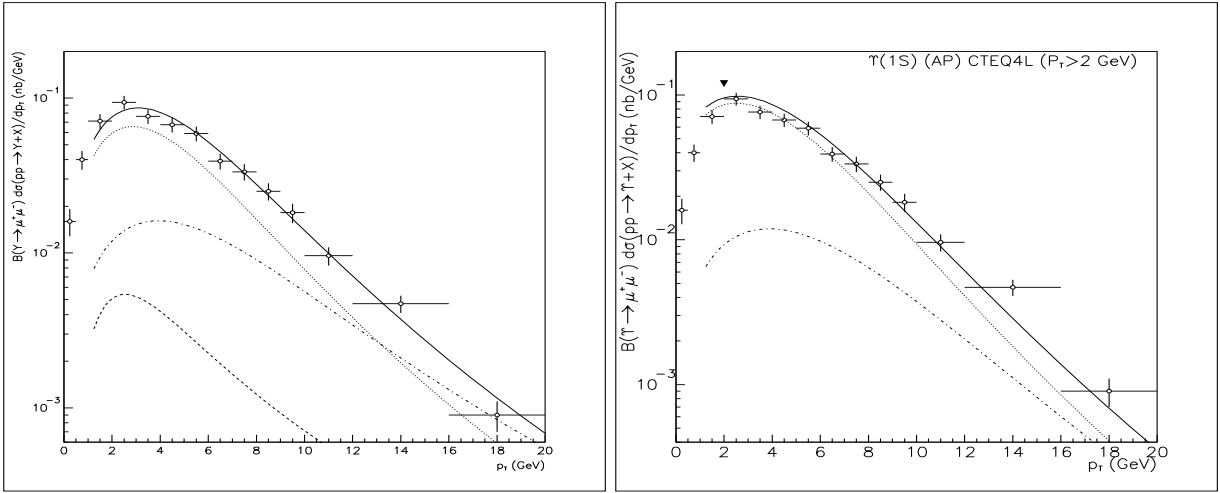,height=6.5cm,width=8.cm}
\psfig{figure=nupsi18_c4l_ap_c2.eps,height=6.5cm,width=8.cm}
}}
\caption{Fits to Tevatron $\Upsilon(1S)$ data using CTEQ2L {\em (left)} 
and CTEQ4L {\em (right)}; 
 dotted line: CSM, dashed line $^1S_0+^3P_J$ contribution, 
dot-dashed line: $^3S_1^{(8)}$ contribution, solid line: all contributions.}
\end{figure}

\subsection*{CTEQ4L versus CTEQ2L}

In our previous work \cite{mas4} on $\Upsilon(1S)$ hadroproduction we 
employed the (now outdated) CTEQ2L parton distribution function, which
however is still often used in current applications
of PYTHIA at LHC collaborations. Throughout 
this  paper we have used CTEQ4L but a comparison
with the previous analysis is in order. In Figure 7 we present the 
two fits to the same Tevatron data \cite{fermi1}. Notice that the 
$^1S_0+^3P_J$ contribution has
been disregarded in the latter case.

\vskip 0.5cm
\begin{table*}[hbt]
\setlength{\tabcolsep}{1.5pc}
\caption{Colour-octet matrix elements (in units of $10^{-3}$ GeV$^3$) from 
the best fits to CDF data at the Tevatron on prompt $\Upsilon(1S)$ 
inclusive production,
using either CTEQ2L [15] or CTEQ4L parton distribution functions 
respectively.}
\label{FACTORES}

\begin{center}
\begin{tabular}{lcc}    \hline
ME:  & $<O_8^{\Upsilon(1S)}(^3S_1)>{\mid}_{tot}$ & 
$M_5^{\Upsilon(1S)}=5{\times}\biggl(\frac{<O_8^{\Upsilon(1S)}(^3P_0)>}{m_b^2}+
\frac{<O_8^{\Upsilon(1S)}(^1S_0)>}{5}\biggr)$ \\
\hline
CTEQ2L & $139{\pm}18$ & $6{\pm}5$ \\
\hline
CTEQ4L & $77{\pm}17$ & ${\simeq}\ 0$  \\
\hline
\end{tabular}
\end{center}
\end{table*}

The CTEQ4L PDF incorporates a BFKL style rise at small $x$, rather than
a flat shape as in CTEQ2L. Therefore it is not surprising that we
find smaller values for the colour-octet matrix elements in the
former case, as can be seen in Table 7.

\newpage

\section{}

\setcounter{equation}{0}

\large{\bf Altarelli-Parisi evolution}

According to the colour-octet model, gluon fragmentation becomes the 
dominant source of heavy quarkonium direct production at high 
transverse momentum. On the other hand, 
Altarelli-Parisi (AP) evolution of the splitting gluon
into ($Q\overline{Q}$)
produces a depletion of its momentum and has to be properly taken
into account. If not so, the resulting long-distance parameter
for the  $^3S_1^{(8)}$ channel would be underestimated from the fit
\cite{montp}.

The key idea is that the AP evolution of the
fragmenting gluon is performed from the evolution of
the {\em gluonic partner} of quarkonium in the final-state
of the production channel

\begin{equation}
g\ +\ g\ {\rightarrow}\ g^{\ast}({\rightarrow} 
(Q\overline{Q})[^3S_1^{(8)}])\ +\ g
\end{equation}
\vskip 0.2cm

Let us remark that, in fact,  $g^{\ast}$ is not
generated in our code \cite{mas2}.  Final hadronization into a 
($Q\overline{Q}$) bound state is taken into
account by means of the colour-octet matrix
elements multiplying the respective
short-distance cross sections \cite{cho,mas2}.
Nevertheless, it is reasonable to assume that, on the average, 
the virtual $g^{\ast}$ should evolve at high $p_T$
similarly to the other final-state gluon - which actually is
evolved by the PYTHIA machinery. 
We used this fact to simulate the (expected) evolution
of the (ungenerated)  $g^{\ast}$ whose momentum was assumed
to coincide with that of the resonance (neglecting the effect
of emission/absorption of soft gluons by the intermediate coloured 
state bleeding off colour \cite{mas3}).
\par
Therefore, event by event we get a correcting factor
to be applied to the transverse mass of the
$(Q\overline{Q})$ state (for the $^3S_1^{(8)}$ channel only):

\begin{equation}
x_{AP}\ =\ \frac{\sqrt{p_T^{{\ast}2}+m_{(Q\overline{Q})}^2}}
{\sqrt{p_T^{2}+m_{(Q\overline{Q})}^2}} 
\end{equation}
where $p_T$ ($p_T^{\ast}$) denotes the transverse momentum of 
the final-state gluon without (with) AP evolution and
$m_{(Q\overline{Q})}$ denotes the mass of the
resonance. At high $p_T$,
\begin{equation}
 p_T^{AP}\ =\ x_{AP}\ {\times}\ p_T
\end{equation}
where $p_T$ is the transverse momentum of the resonance
as generated by PYTHIA (i.e. without AP evolution), whereas
for $p_T\ {\leq}\ m_{(Q\overline{Q})}$ the effect becomes
much less significant as it should be. Thus the interpolation between
low and high $p_T$ is smooth with the right asymptotic
limits at both regimes.
The above way to implement AP evolution may appear 
somewhat simple but it remains in the spirit of our whole
analysis, i.e. using PYTHIA machinery whenever possible. In fact, 
it provides an energy depletion of the fragmenting gluon
in agreement with previous work on
charmonium hadroproduction \cite{cho,montp}.
In Figure 8 the $x_{AP}$ factor is plotted as a function of the transverse 
momentum of the resonance for the Tevatron event generation. 
\vskip 0.5cm
\begin{figure}[htb]
\centerline{
\psfig{figure=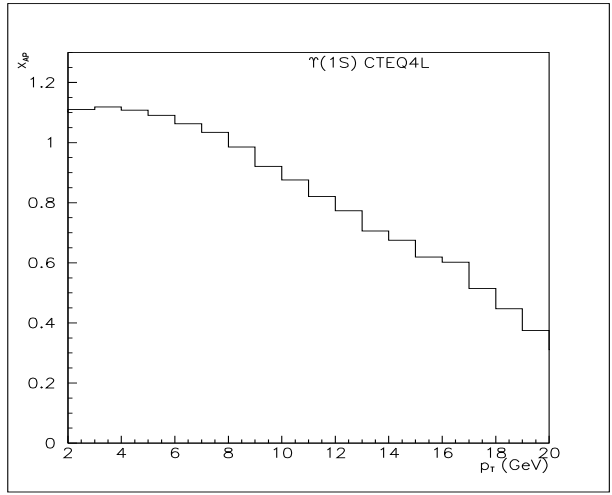,height=6.5cm,width=8.cm}
}
\caption{$x_{AP}$ factor as a function of $p_T$ for Tevatron energies
obtained from our generation.}
\end{figure}

Moreover, in order to assess the effect of AP evolution on the fit parameters
we show in Table 8 two numerical values for the relevant
colour-octet MEs obtained from a best ${\chi}^2$
fit to Tevatron data \cite{fermi1} using the CTEQ4L PDF: (i) the first
row corresponds to a generation {\em without} AP evolution; (ii)
the second one does take it into account. Notice the
increase of $<O_8^{\Upsilon(1S)}(^3S_1)>{\mid}_{tot}$  
in the latter case w.r.t.
AP off, but to a lesser extent than for charmonium \cite{montp}.

\vskip 0.5cm
\begin{figure}[htb]
\centerline{\hbox{
\psfig{figure=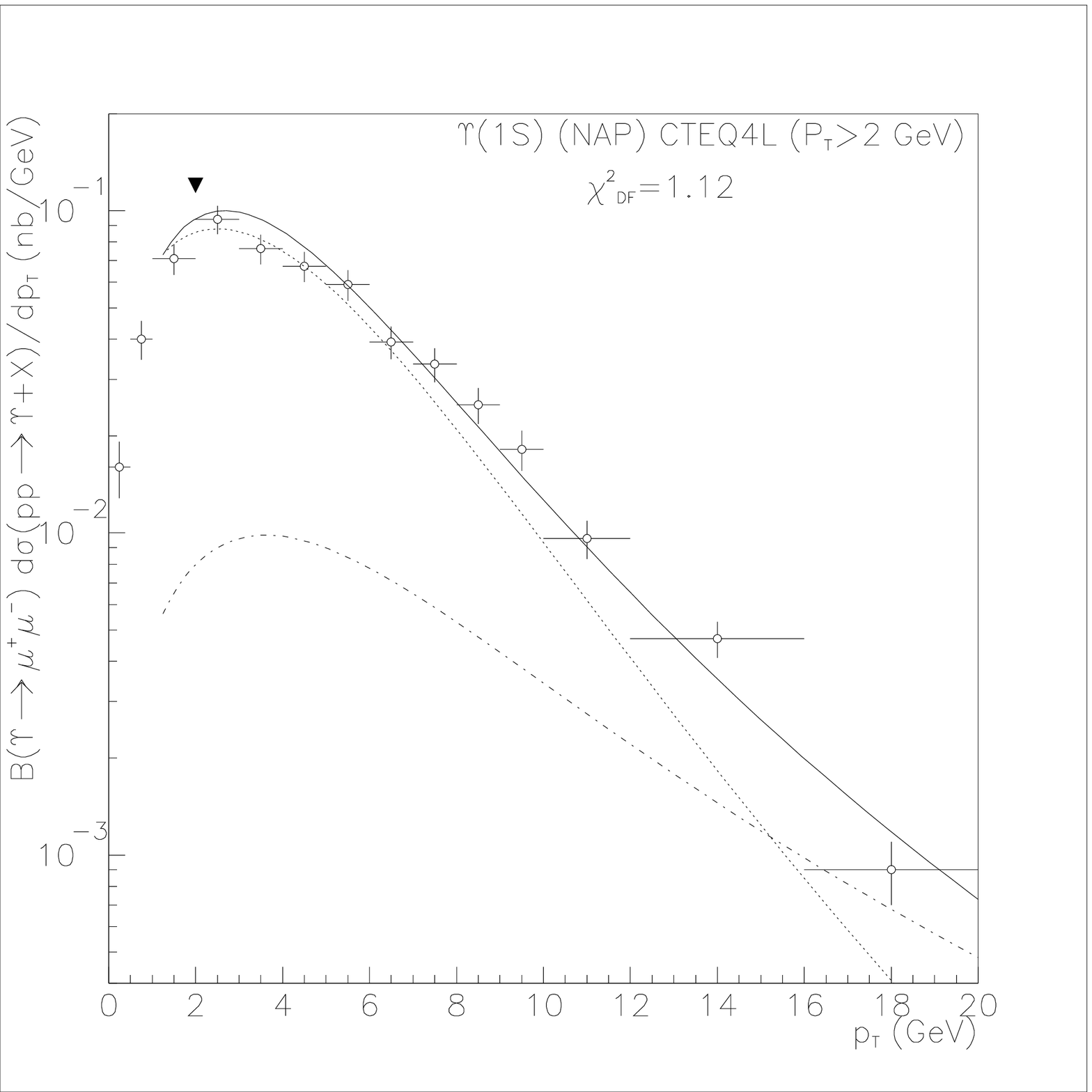,height=6.5cm,width=8.cm}
\psfig{figure=nupsi18_c4l_ap_c2.eps,height=6.5cm,width=8.cm}
}}
\caption{Theoretical curves obtained from a fit using PYTHIA including 
the colour-octet mechanism for prompt $\Upsilon(1S)$ production against CDF
data at the Tevatron
{\it a)} without AP  evolution of the fragmenting gluon, {\it b)} 
with AP evolution of the fragmenting gluon. The  CTEQ4L parton distribution
function and $m_b=4.88$ GeV were employed in the fits; dotted line: CSM,  
dot-dashed line: $^3S_1^{(8)}$ contribution, solid line: all contributions.} 
\end{figure}
\vskip 0.5cm

\begin{table*}[hbt]
\setlength{\tabcolsep}{1.5pc}
\caption{Colour-octet matrix elements (in units of $10^{-3}$ GeV$^3$) from 
the best fit to CDF data at the Tevatron on prompt $\Upsilon(1S)$ production.
The CTEQ4L PDF was used with AP
evolution off and on respectively.}
\label{FACTORES}

\begin{center}
\begin{tabular}{lc}    \hline
ME:  & $<O_8^{\Upsilon(1S)}(^3S_1)>{\mid}_{tot}$  \\
\hline
AP off & $70{\pm}15$   \\
\hline
AP on & $77{\pm}13$   \\
\hline
\end{tabular}
\end{center}
\end{table*}
\vskip 0.5 cm

It is worth noting that the effect of the AP evolution
on the shape of the differential cross section
over the [1,20] GeV $p_T$-range, though sizeable, 
is considerably less pronounced for bottomonium than
for charmonium \cite{montp} because of the larger mass of the former.
Nevertheless we can appreciate in Figure 9 that the plot corresponding to
AP evolution is noticeably steeper at moderate and high $p_T$
as could be expected. 
Let us finally remark that, although we can switch on/off AP evolution and
initial-state radiation {\em at will} in the event generation, both 
next-to-leading order effects have to be incorporated for a realistic
description of the hadronic dynamics of the process.

\newpage

\section{}

\large{\bf Rapidity cut and azimuthal correlations}

\setcounter{equation}{0}

In this appendix we show that the (systematic) uncertainty 
associated to the determination of the Feynman $x$ 
of the interacting partons in our proposed method, 
is given by the upper rapidity cut $y_0$ on the
resonance imposed to events, according to
the expression:

\begin{equation}
\frac{{\Delta}x}{x}\ =\ y_0
\end{equation}

Indeed, assuming a gluon gluon scattering process into two 
final-state gluons, it is 
easy to see that any extra (longitudinal) rapidity amount
${\Delta}y$ of any final-state parton, should be assigned to anyone of 
the two colliding partons, as a consequence of conservation of 
energy-momentum. (In this case the partonic reference frame would
not longer coincide with the Lab frame.) 

On the other hand, a parton 
carrying a fraction $x$ of the total hadron momentum has a 
(longitudinal) rapidity

\begin{equation}
y\ =\ y_{hadron}\ -\ \log{\frac{1}{x}}
\end{equation}

where $y_{hadron}$ is the rapidity of the hadron in the Lab system.

Differentiating both sides of Eq. (C.2) and setting ${\Delta}y=y_0$, one
gets easily the expression (C.1).

Let us observe that the rapidity cut 
${\mid}y{\mid}<y_0$ binds us to a region of $\lq\lq$allowed'' 
transverse momentum, increasing with $p_T$ since $x^2s\ {\simeq}\ 4p_T^2$, and 
hence 

\begin{equation}
\frac{{\Delta}p_T}{p_T}\ =\ y_0
\end{equation}

This means that as the transverse momentum grows, the $p_T$
range compatible with the relative error, predetermined by
choosing the value of $y_0$, grows too. If this value is set very low,
the precision on the Feynman $x$ increases but the price to be paid
is probably reducing
too much the statistics. Conversely, allowing $y_0$ to be 
{\em too large}, leads to
larger statistics but spoiling the knowledge of $x$ because of
the uncertainty given by (C.1). As a
compromise, we chose $y_0=0.25$ which, however, could be varied
depending on the size of the sample of collected events.

\begin{figure}
\centerline{\hbox{ 
 \psfig{file=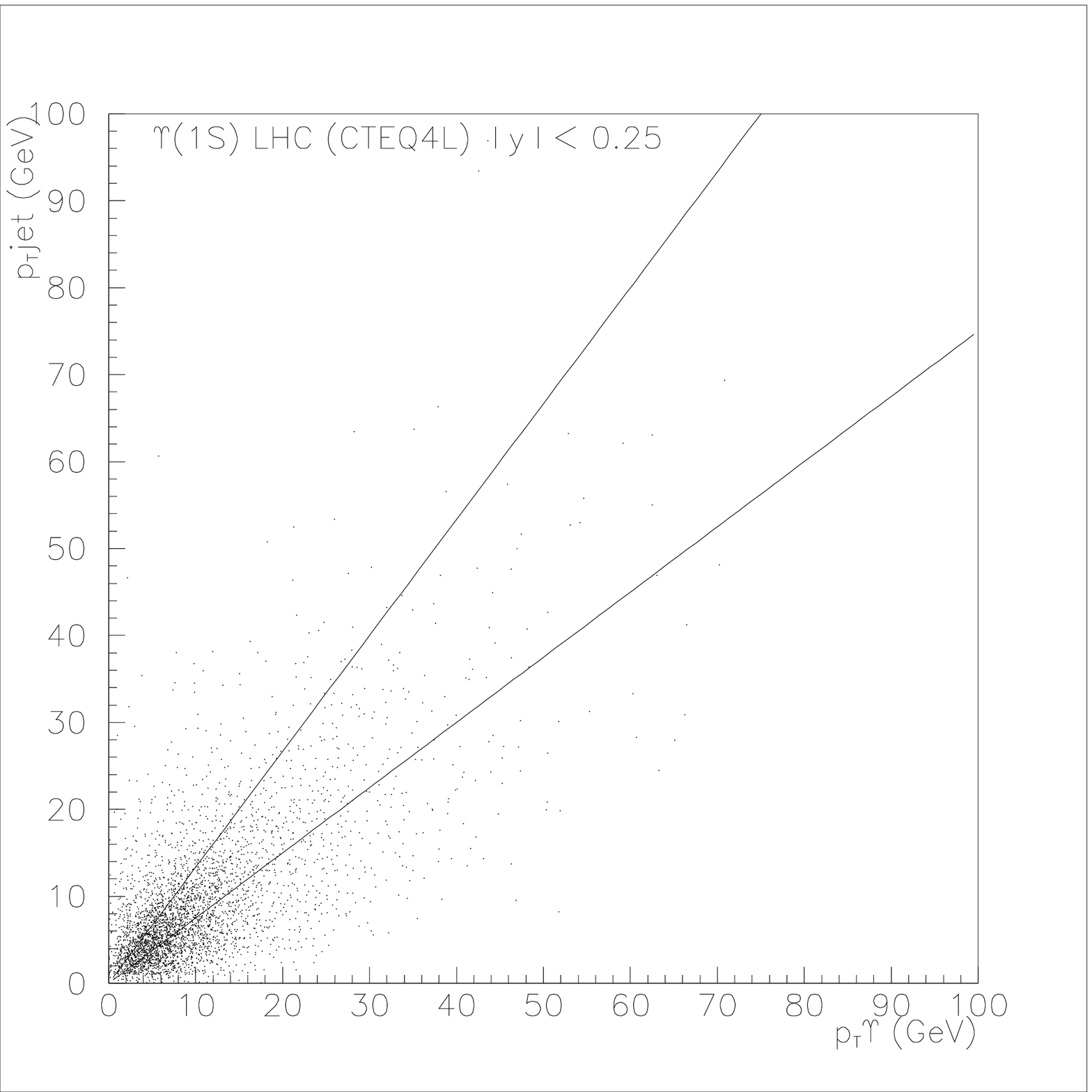,width=5.5cm} 
 \psfig{file=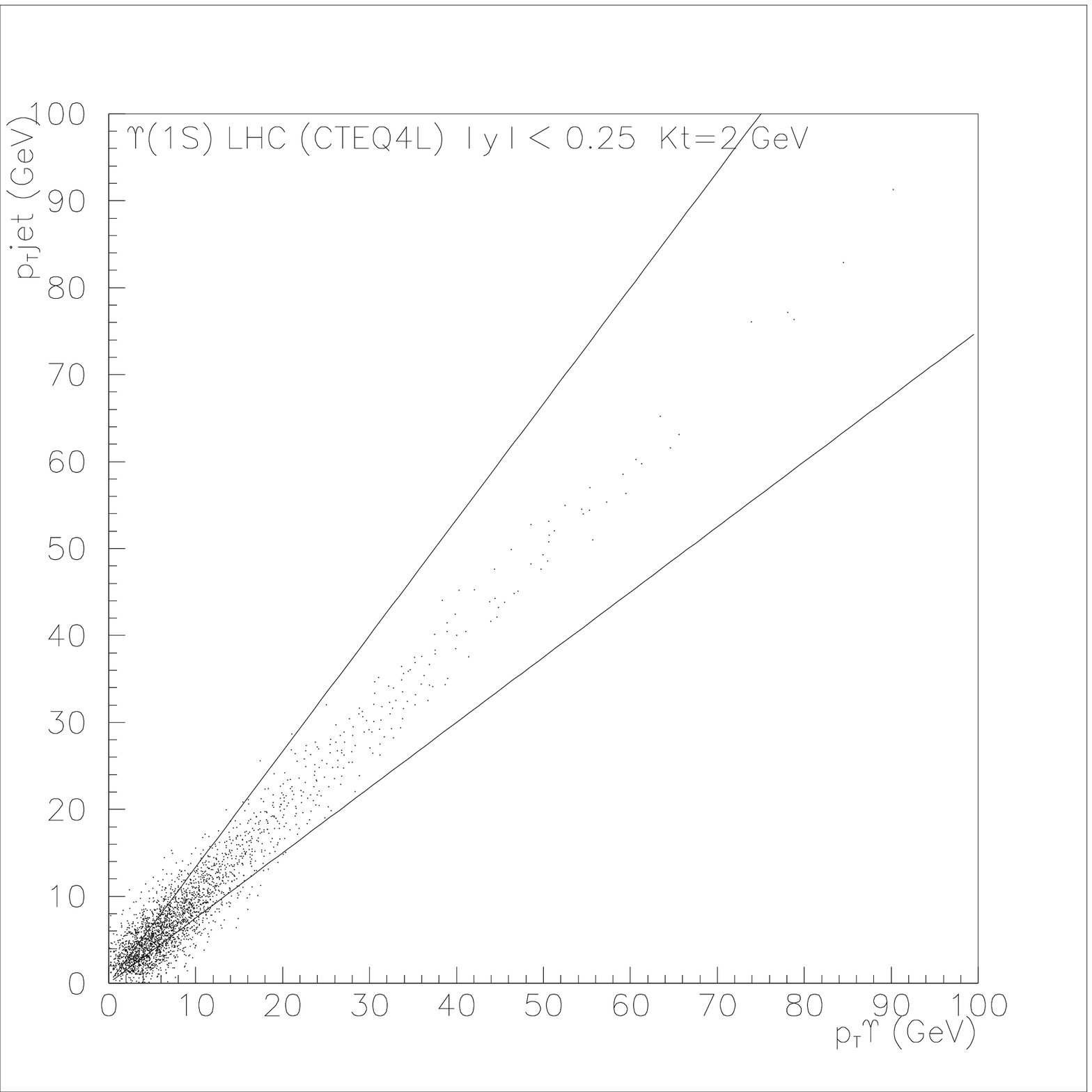,width=5.5cm} 
 \psfig{file=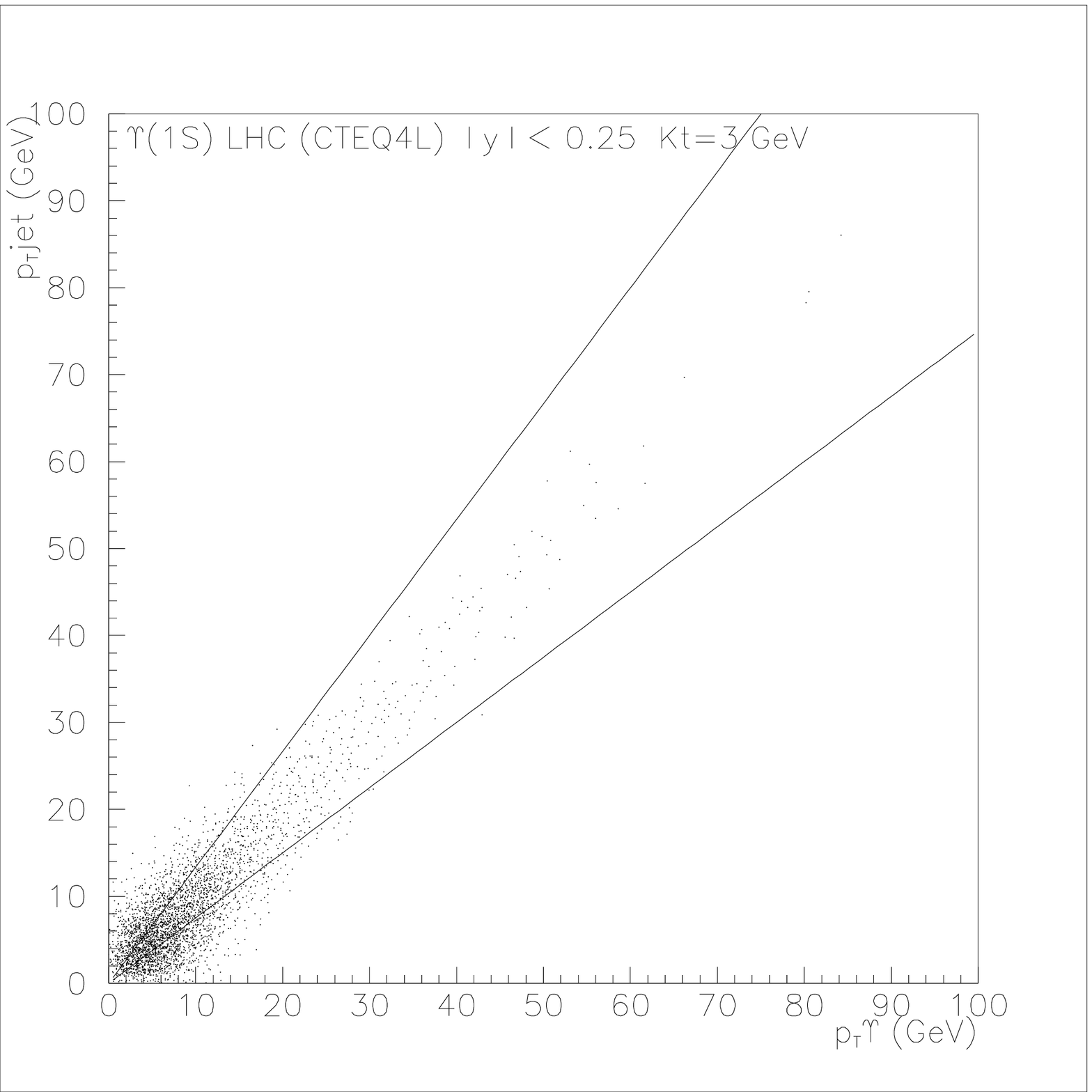,width=5.5cm}  
}}
\caption{Plots of the jet transverse momentum versus the
$\Upsilon(1S)$ resonance 
transverse momentum at LHC energy (parton/particle level simulation)
using, from left to right: {\em a)} the PYTHIA algorithm
to simulate initial-state radiation; using a gaussian
smearing function with {\em b)} $<k_T>=2$ GeV and {\em c)} $<k_T>=3$ GeV.
The two straight lines indicate the allowed region according
to the $p_T$ uncertainty obtained from Eq. C.3 for $y_0=0.25$.}
\end{figure}

In order to get an idea of the expected impact of the 
intrinsic $k_T$ on the topology of events, we show in Figure 10 several
plots of the transverse momenta of the $\Upsilon(1S)$
resonance versus the recoiling jet. In the absence of any
higher order QCD effect, events squeeze along the
diagonal. However $k_T$ smearing spreads events over a larger area
in the plot, spoiling somehow a naive picture of a back-to-back topology
coming from a collinear approximation to leading order; 
Figure 10.a) corresponds to initial-state radiation
activated in the PYTHIA generation following the
model developed in \cite{torn2}. Alternatively, 
Figures 10.b) and 10.c) show the effect
of a gaussian spread of
$<k_T>=2$ GeV and $<k_T>=3$ GeV, respectively. The region
inside the two straight lines corresponds to an uncertainty on
$p_T$ given by Eq. (C.3) for a rapidity value $y_0=0.25$.
Although at small and moderate $p_T$ (say, $p_T\ {\leq}\ 10$ GeV)  
all plots essentially agree, at higher $p_T$ the
former one, corresponding to a full simulation
of gluon emission in the initial-state performed
by PYTHIA, displays much more events outside the accepted
region.
\vskip 0.5cm

\begin{table} [htb]
\caption{Fraction (in $\%$) of events 
inside the region defined by the two straight lines for  
different $p_T$ lower cuts (in GeV) applied to the resonance,
corresponding to Fig. 10.a), i.e. initial-radiation generated by
PYTHIA.}
\begin{center}
\begin{tabular}{|c|c|c|c|c|c|}
\hline
$p_T$ cut-off: & $10$ & $20$ & $30$ & $40$  & $50$ \\
\hline
$\%\ \lq\lq$inside'' & $39 \pm 1$ & $38 \pm 3$ & $35 \pm 5$ & $38 \pm 9$ & $38 \pm 15 $\\
\hline

\end{tabular}
\end{center}
\end{table}

In Table 9 we show the fractions of events inside the allowed region
between the two straight lines in the plot 10.a) (initial-state
radiation on). We observe that about $40\%$ of all events are 
$\lq\lq$accepted'',   
remaining practically constant above $p_T=10$ GeV. Finally
we conclude that such reduction factor (of the order of
$40\%$) does not represent in itself a dramatic loss of statistics 
regarding our proposed method to probe the gluon density in protons.
On the other hand, for the gaussian smearing, the situation
is even much more optimistic.

\begin{figure}[htb]
\centerline{\hbox{ 
 \psfig{file=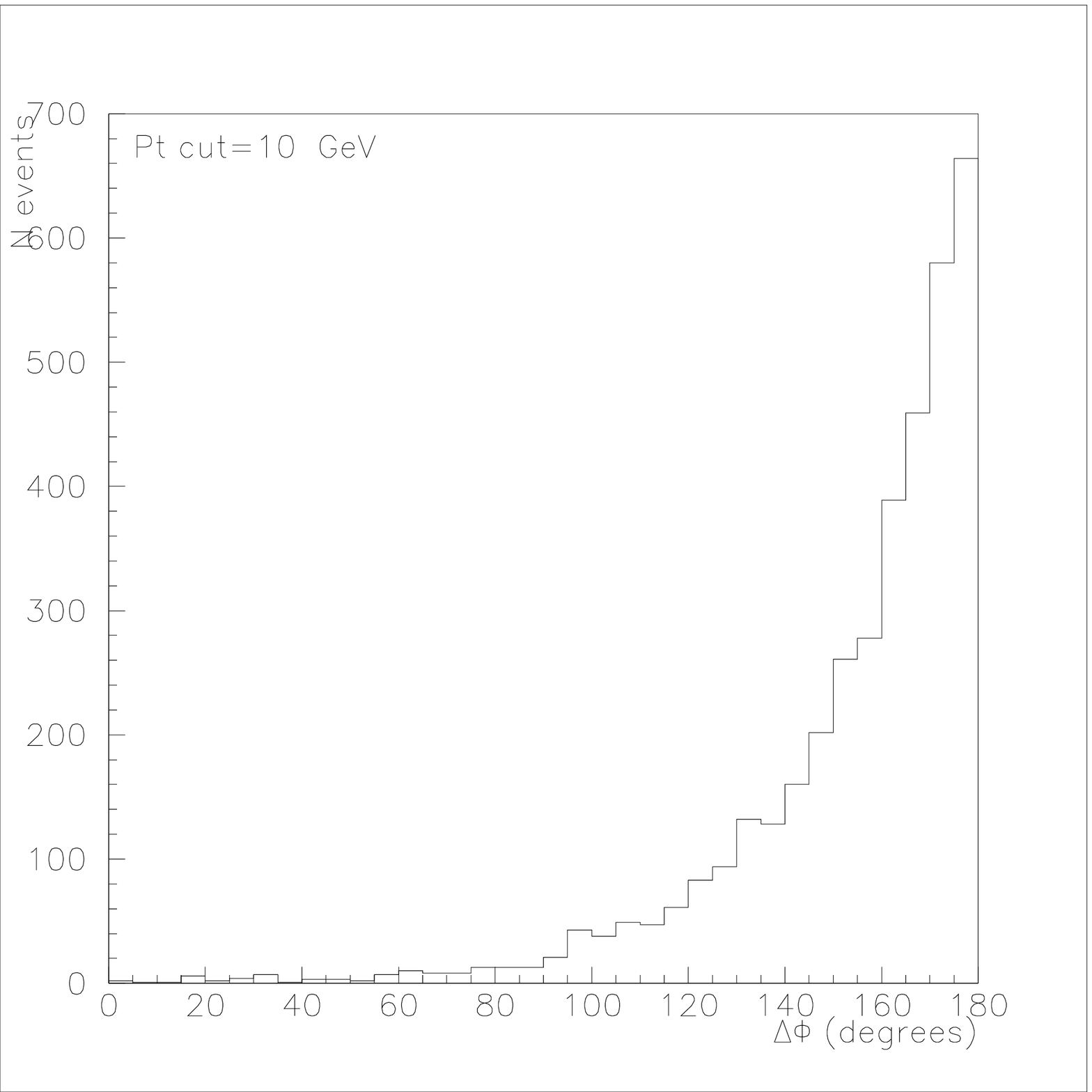,width=5.5cm}
 \psfig{file=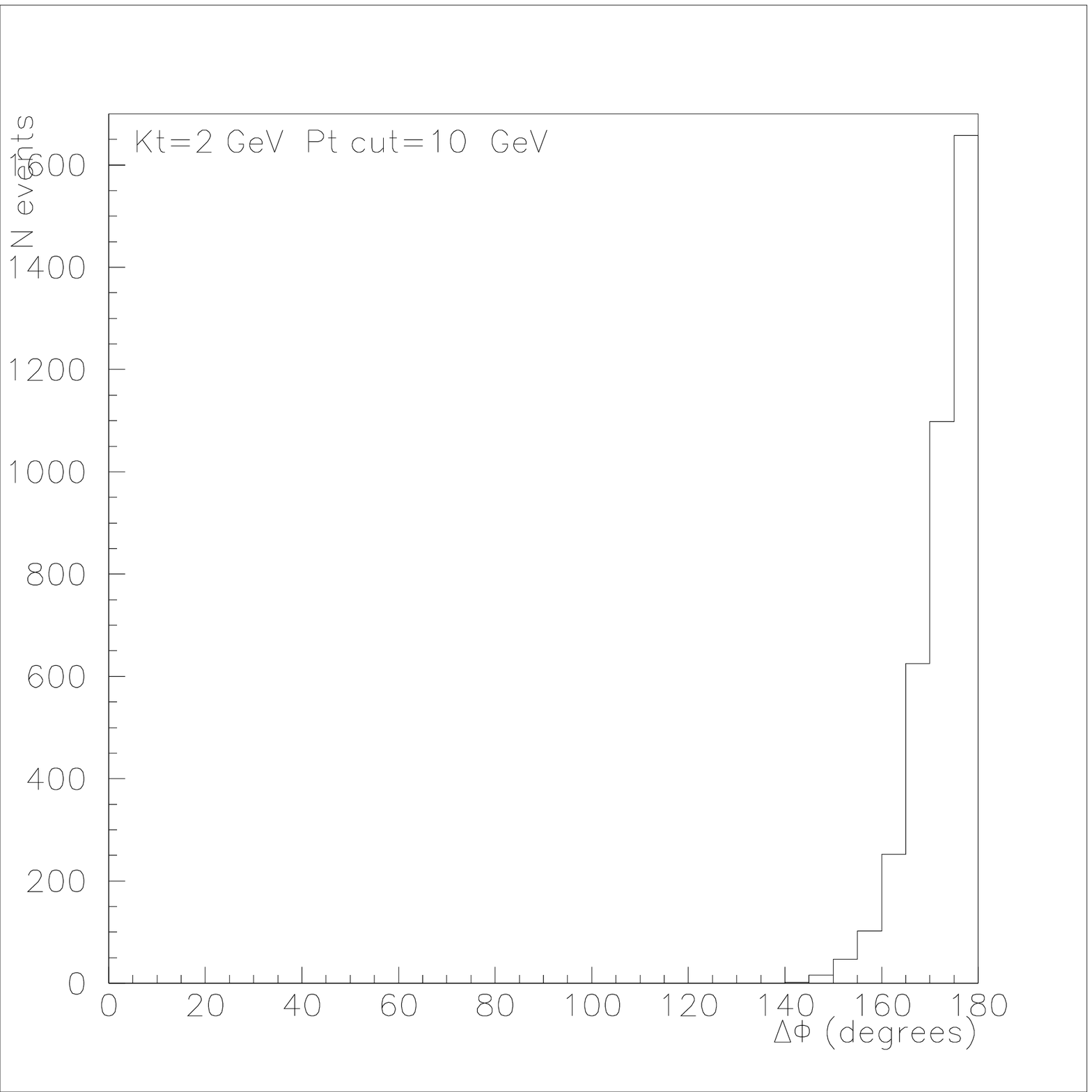,width=5.5cm}
 \psfig{file=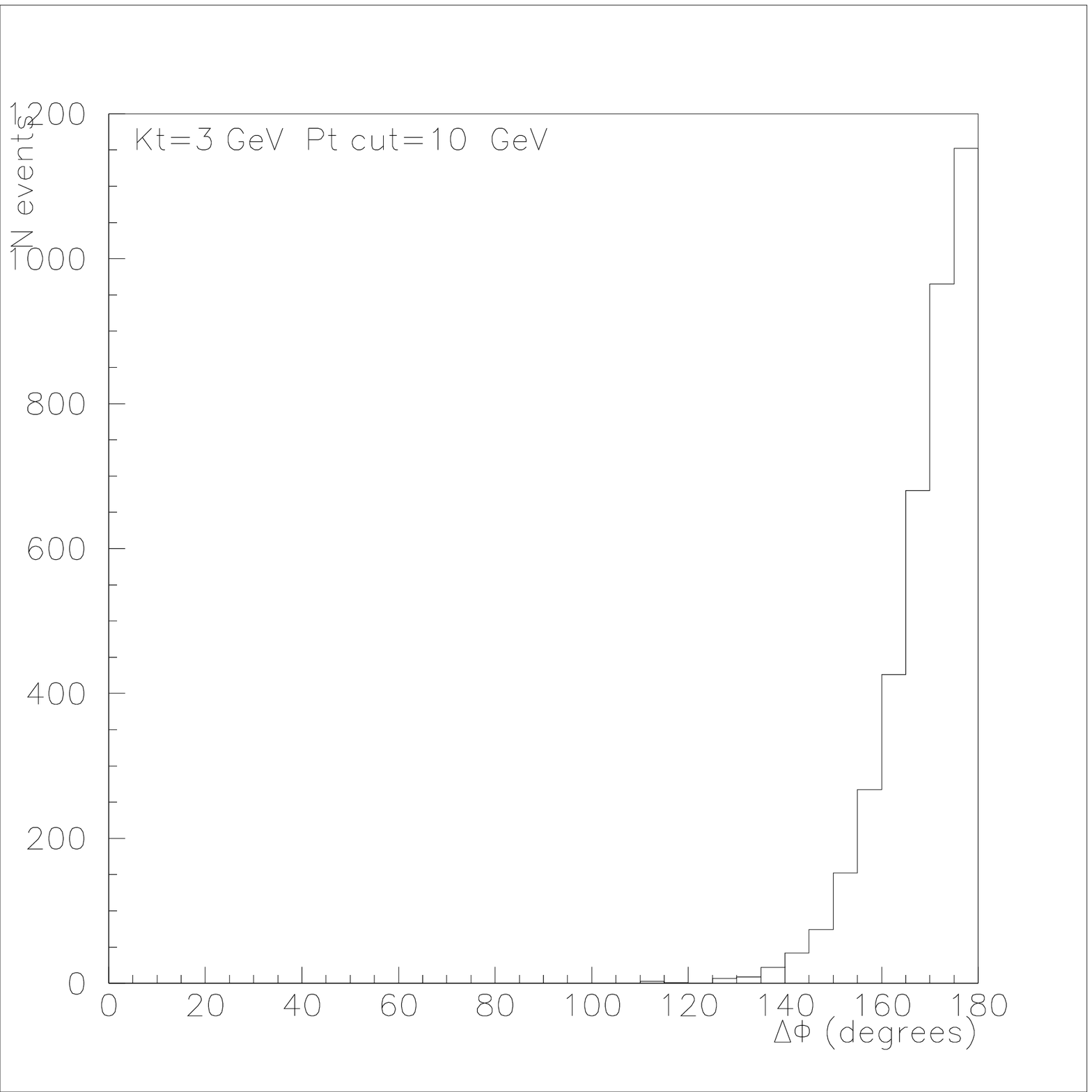,width=5.5cm} 
}}
\caption{Azimuthal angle between the recoiling jet direction
(defined by the parent gluon momentum) 
and the dimuon direction from $\Upsilon(1S)$ decays in the 
transverse plane, from left to right:
{\em a)}: Initial-state radiation activated in PYTHIA; {\em b)} Using
gaussian smearing with $<k_T>=2$ GeV; {\em c)} The same
with $<k_T>=3$ GeV. All plotted events were selected
with $\Upsilon(1S)$ transverse momentum greater than 10 GeV.
}
\end{figure}

\vskip 0.5cm

In Figures 11 we show the azimuthal $\Delta{\phi}$ angle between the
muon pair direction (defining the direction of the fragmenting
gluon into bottomonium) and the recoiling jet generated by the 
final-state gluon, for different values of the
effective $k_T$, in correspondence with Figures 10. In Fig. 11.a)
we used the PYTHIA algorithm for initial-state radiation, whereas
in Figures 11.b) and 11.c) we used a smearing gaussian with
$<k_T>=2$ GeV and $<k_T>=3$ GeV, respectively. As expected,
again we realize the sizeable effect
of the effective $k_T$ effect on the distribution, 
especially in the former case. Nevertheless, most
events should display a clear enough back-to-back signature
as regards the $\Delta{\phi}$  variable
(in addition to the $p_T$ balance), as indicated by 
the peak at 180 degrees in all plots of Figure 11.

\end{document}